\documentclass[12pt,preprint]{aastex}

\shorttitle{On the Soft Excess in the X-ray Spectrum of Cir X--1}
\shortauthors{Iaria et al.}

\begin{document}
 
\title{On the Soft Excess in the X-ray Spectrum of Cir X--1: Revisitation
 of the Distance to Cir X--1}

\author{R. Iaria\altaffilmark{1}, M.  Span\`o\altaffilmark{1}, T. Di Salvo\altaffilmark{1}, N. R.
  Robba\altaffilmark{1}, L. Burderi\altaffilmark{1,2}, 
 R. Fender\altaffilmark{3},
M. van der Klis\altaffilmark{3}, F. Frontera \altaffilmark{4,5}}

 \altaffiltext{1}{Dipartimento di
 Scienze Fisiche ed Astronomiche, Universit\`a di Palermo, Via
 Archirafi 36, 90123 Palermo, Italy     
              email: iaria@gifco.fisica.unipa.it}
        
 \altaffiltext{2}{Osservatorio Astronomico di Roma, Via Frascati 33,
 00040 Monteporzio Catone (Roma), Italy   }

 \altaffiltext{3}{Astronomical Institute "Anton Pannekoek," University of
Amsterdam and Center for High-Energy Astrophysics,
Kruislaan 403, NL 1098 SJ Amsterdam, the Netherlands.
}
 
 \altaffiltext{4}{Dipartimento di Fisica, Universita' degli Studi di Ferrara, 
Via Paradiso 12, I-44100 Ferrara, Italy}

 \altaffiltext{5}{Istituto Astrofisica Spaziale e Fisica Cosmica, CNR, 
Via Gobetti 101, I-40129 Bologna, Italy}

\begin{abstract}
  
  We  report on  a  300  ks BeppoSAX  (0.12--200  keV) observation  of
  Circinus X--1 (Cir X--1) at phases between 0.62 and 0.84 and on a 90
  ks BeppoSAX observation of Cir  X--1 at phases 0.11--0.16. Using the
  canonical model adopted until now to fit the energy spectrum of this
  source large  residuals appear below  1 keV.  These are  well fitted
  using  an  equivalent  hydrogen  column  of  $0.66  \times  10^{22}$
  cm$^{-2}$,  adding absorption  edges of  \ion{O}{7},  \ion{O}{8} and
  \ion{Ne}{9} in the spectra  extracted from the observation at phases
  0.62--0.84  and adding absorption  edges of  \ion{O}{7}, \ion{O}{8},
  \ion{Mg}{11} and \ion{Mg}{12} and absorption lines of \ion{O}{8} and
  \ion{Mg}{12} in the spectra extracted from the observation at phases
  0.11--0.16.   During  the   observation  at  phases  0.62--0.84  the
  electron density associated to  the ionized matter is $\sim 10^{13}$
  cm$^{-3}$  remaining  quite constant  going  away  from the  compact
  object.  During  the observation  at phases 0.11--0.16  the electron
  density profile varies along the  distance going from $\sim 6 \times
  10^{13}$ cm$^{-3}$ at  $\sim 10^{11}$ cm to $\sim  9 \times 10^{10}$
  cm$^{-3}$  at $\sim  10^{13}$  cm.  The  equivalent hydrogen  column
  towards Cir X--1  is three times lower than  the value obtained from
  previous models.  This  low value would imply that Cir  X--1 is at a
  distance of 4.1 kpc.

\end{abstract}


\maketitle

\section{Introduction}

Cir X--1  is a peculiar Low  Mass X-ray Binary  (LMXB), which recently
showed ultra-relativistic outflows similar to those produced by active
galactic  nuclei   (Fender  et  al.   2004).  Because  of   its  rapid
variability, a  black hole  was supposed to  be the compact  object in
this  system (Toor  1977)  until  type I  X-ray  bursts were  detected
(Tennant et al. 1986a, 1986b), suggesting that the compact object is a
neutron star.

Until 2002 Cir X--1 was supposed to be a runaway binary because of its
supposed association  with the supernova remnant  G321.9-0.3 (Clark et
al.   1975)  at a  distance  between  5.5 kpc  and  6.5  kpc (Case  \&
Bhattacharya,  1998; Stewart  et  al.  1993).   However,  the work  of
Mignani et  al.  (2002), using  HST data, excluded the  association of
Cir  X--1 with  G321.9-0.3.  This  brings to  a new  debate  about the
distance to Cir  X--1. From the observed 21-cm  absorption features in
the radio  spectrum of  the source  a lower limit  to the  distance of
6.4--8~kpc  was  obtained  (Goss   \&  Mebold,  1977;  Glass,  1994).  
Generally in LMXBs  the companion star has a  mass $\la 1$ M$_{\odot}$
and the  orbit is almost  circular; contrarily the  canonical scenario
accepted for Cir  X--1 argues that the source has  a companion star of
$\sim  3-5$ M$_{\odot}$  (probably a  subgiant,  see Johnston  et al.  
1999), an orbital  period of $\sim 16.6$ days,  deduced from the radio
and x-ray lightcurve of the source, and an eccentric orbit with $e\sim
0.7-0.9$ (Murdin et al.  1980; Tauris et al. 1999).

The correlated  spectral and  fast-timing properties of  accreting low
magnetic neutron  stars allow distinguishing them into  two classes, Z
and atoll sources, using the pattern each source describes in an X-ray
color-color diagram (Hasinger \& van der Klis 1989). The Z sources are
very  bright and  could have  a moderate  magnetic field  ($< 10^{10}$
Gauss), while the atoll sources have a lower luminosity and probably a
weaker magnetic  field.  The timing  analysis of RXTE data  (Shirey et
al.  1996) evidenced  that Cir X--1 is more similar to  LMXBs of the Z
class.    However  no   quasi  periodic   oscillations   at  kilohertz
frequencies were observed in Cir X--1 (Shirey, 1996) contrarily to all
the other Z-sources.

Cir X--1 also shows a  peculiar X-ray spectrum.  Brandt et al. (1996),
using ASCA/GIS data taken in August 1994, found a sudden variation (in
a timescale  of $\sim  20$ minutes)  of the flux  at phase  just after
zero, where the count rate increased from $\sim 30$ counts s$^{-1}$ to
$\sim 300$ counts s$^{-1}$. In  the low-count rate state they obtained
a good  fit to the  0.6--10 keV spectrum  of Cir X--1 using  a partial
covering  component, in  addition  to a  two-blackbody  model, with  a
corresponding  hydrogen  column of  $\sim  10^{24}$  cm$^{-2}$ and  an
intrinsic absorption  of $\sim 1.7  \times 10^{22}$ cm$^{-2}$;  in the
high-count rate state  the partial covering of the  X-ray spectrum was
not needed.  Because  of the similarity with the  spectra of Seyfert 2
galaxies with Compton-thin tori,  Brandt et al.  (1996) suggested that
matter  at the  outer edge  of the  accretion disk,  together  with an
edge-on disc  orientation, could explain  the partial covering  of the
spectrum of Cir X--1 (see, however, Fender et al. 2004).

Iaria et al. (2001a, 2002), using BeppoSAX data, analysed the 0.1--100
keV spectrum  of Cir X--1  at phases 0.11--0.16 and  0.61--0.63.  They
found that the  continuum could be well described  by a Comptonization
model at phases 0.11--0.16 and  by the combination of a blackbody plus
a Comptonization  model at phases 0.61--0.63;  the blackbody component
is probably produced in the  inner region of the accretion disc, while
the Comptonized component is produced by a (hotter) corona surrounding
the neutron star.  A hard excess, dominanting the spectrum at energies
higher than  15 keV,  was also observed  in both the  observations and
fitted  by  a power-law  or  a  bremsstrahlung  component.  A  similar
feature was  observed in  other Z-sources  (GX 17+2, Di  Salvo et  al. 
2000; GX 349+2, Di  Salvo et al. 2001; GX 5--1, Asai  et al. 1994; Sco
X--1, D'Amico et al. 2000; Cyg X--2, Di Salvo et al. 2002), suggesting
that Cir  X--1 shows a  behavior similar to  that of other  Z-sources. 
Finally a  strong absorption edge  at $\sim 8.4-8.7$ keV,  produced by
highly ionized  iron, was needed  to fit the  spectrum of Cir  X--1 at
phases 0.11--0.16.

The presence of  a hard power-law component was  recently confirmed by
RXTE (3--200 keV)  observations of Cir X--1 (Ding  et al. 2003).  They
extracted the source spectra at different positions in the color-color
diagram  and concluded  that there  is no  clear relation  between the
position of the source in  the color-color diagram and the presence of
the hard tail in the energy spectrum.  An absorption edge, with energy
ranging from 8.3  keV to 9.2 keV, was again  observed, together with a
localized emission feature present around 10 keV in the energy spectra
corresponding  to the  normal  branch and  flaring  branch; a  similar
feature was observed by Shirey et al. (1999). They interpret this line
as produced by heavy elements, such as nickel, or, less probably, as a
blue-shifted iron line  due to the motion in a  relativistic jet or to
the rotation  of the accretion disk; however  extreme conditions would
be required to boost the line energy up to 10 keV in this case.

Iaria et al. (2001b), using  ASCA data, studied the energy spectrum of
Cir  X--1 along its  orbit. They  distinguished three  different X-ray
states of  the source as  a function of  its phase.  Therefore  it was
possible to study the evolution of the absorption edge along the orbit
of Cir  X--1. At  the phase  of the flaring  activity, near  the phase
zero, the  hydrogen column  derived from the  edge was  $\sim 10^{24}$
cm$^{-2}$;  at  larger  values   of  the  phase  the  hydrogen  column
decreased;  the absorption  edge  (from ionized  iron)  was no  longer
detected at  phases from 0.78  to 1, while  at these phases  a partial
covering component was required  with an equivalent hydrogen column of
$\sim 10^{24}$ cm$^{-2}$.

In general  the X-ray spectrum of  Cir X--1 is quite  rich of discrete
features.  Brandt  \& Schulz (2000), using Chandra  data, observed Cir
X--1 near the phase zero. They found P-Cygni profiles of lines emitted
by highly ionized elements.  The outflow velocity necessary to explain
the P-Cygni profiles was $\sim  2000$ km/s.  Under the hypothesis that
the source is  seen nearly edge-on, they supposed that  the model of a
thermally  driven  wind (Begelman  et  al.  1983)  could explain  this
outflow of  ionized matter from  the system. Also they  suggested that
the  absorption  edge  observed  by  Iaria et  al.  (2001a)  could  be
connected to the outflow.
   
The main  point that  should be  clarified is the  cycle of  16.6 days
observed in the radio and X-ray lightcurve that was supposed to be the
orbital period of  the source. Just before the  phase zero, associated
to  the  radio  outbursts,  a  dip  is present  in  the  x-ray  folded
lightcurve that was  associated to the eclipse of  the emitting source
by  the  companion  star.   The  long-term variability  of  the  x-ray
lightcurve  and the  large flaring  activity, present  near  the phase
zero, carried to the conclusion  that Cir X--1 has an elliptical orbit
with a large value of its  eccentricity (Murdin et al., 1980) and that
the phase zero  corresponds to the periastron of  the orbit.  As noted
by  Clarkson et  al.  (2004)  the  dips are  near-coincident with  the
presumed periastron passage suggesting  that the semimajor axis of the
binary orbit would have to be almost aligned with our line of sight to
the binary system requiring  an improbable fine tuning.  Saz Parkinson
et al.  (2003), studying the RXTE/ASM lightcurve of Cir X--1 noted that
the periodicity  of 16.6 days  is not always observed.   They analysed
the data from 6 Jan 1996 to 19 Sep 2002 finding that the cycle of 16.6
day is not present in a temporal interval lasting 270 days and that in
this interval a  new weak periodicity of 40 days  is observed.  If the
cycle of 16.6  days is the orbital  period of Cir X--1 is  not easy to
explain why  during around  16 orbital periods  the periodicity  is no
more observed  while a new periodicity  of 40 days  appears.  Then the
results reported by Saz Parkinson et al.  (2003)
and Clarkson et  al.  (2004) seem to indicate that  the cycle of 16.6
days in  not connected  to the  orbital period of  the binary  system. 
Recently, from three radio  observations Fender et al.  (2004) resolve
the structure  of a jet  in Cir  X--1. The jet  is extended up  to 2.5
arcsec from the core.  The  authors, assuming a distance to the source
of 6.5 kpc,  find a superluminal motion of the  jet having an apparent
velocity of $v_{app}  \simeq 15 c$; because of the  large value of the
apparent  velocity  the  angle  between  the line  of  sight  and  the
direction  of the  jet, $\theta$,  is $  < 5^\circ$  and  the intrinsic
velocity  of  the  jet is  $\beta  >  0.998$.   The detection  of  the
superluminal  motion in the  jet of  Cir X--1  implies that  $\theta <
5^\circ$   and  since  the   jet  should   have  a   direction  almost
perpendicular to the accretion disk it implies that Cir X--1 could not 
be an edge-on source contrarily to  what was  supposed until  now. 
 
In this  work we assume  the ephemeris as  reported by Stewart et  al. 
(1991).   We present the  results of  the analysis  of the  broad band
(0.12--200 keV) spectra of Cir  X--1 from a BeppoSAX observation taken
at phases 0.62--0.84 in September 2001 and the reanalysis of the broad
band spectrum of  Cir X--1 from a previous  BeppoSAX observation taken
at phases  0.11--0.16 in  August 1998 (see  Iaria et al.,  2001a).  We
note that, although the used model in that paper well fitted the data
in the broad  energy band 0.1--200 keV, a small  bump in the residuals
below 1 keV was present (see Fig.  3 in Iaria et al., 2001a).  We will
show that to well  fit the data below 1 keV we  need to add absorption
edges  of  H-like and  He-like  ions  of  oxygen, neon  and  magnesium
confirming the  presence of highly  ionized matter around the  system. 
Moreover we discuss  the possibility that Cir X--1  is distant 4.1 kpc
implying  that  it  does  not  show a  super-Eddington  luminosity  as
supposed until now.

The  BeppoSAX observation of  Cir X--1  at phases  0.61--0.63, already
published by Iaria et al. (2002)  and cited below in the text, will be
not  reanalysed because the  available energy  range for  the spectral
study is 1.8--200  keV, not allowing us to study  the spectrum below 1
keV.

\section{Observations}
Two pointed observations of Cir  X--1 were carried out in 2001 between
Sep 08 00:47:57 UT and Sep 11 07:34:26 UT and in 1998 Aug 1 15:59:54.5
UT (this latter  observation  was already  analyzed  by Iaria  et al.,  
2001a).  The total exposure time  was, respectively, 115 ks and 90 ks
with the Narrow Field Instruments,  NFIs, on board BeppoSAX (Boella et
al.  1997).  These consist of four co-aligned instruments covering the
0.1--200~keV  energy range:  a  Low-Energy Concentration  Spectrometer
(LECS;  operating   in  the  range   0.1--10~keV),  two  Medium-Energy
Concentration Spectrometers  (MECS; 1.3--10~keV), a  High-Pressure Gas
Scintillation Proportional Counter (HPGSPC; 7--60~keV), and a Phoswich
Detector System (PDS; 13--200~keV). We selected data from the LECS and
MECS images in  circular regions centered on the  source with $8'$ and
$4'$  radius,  respectively. Data  extracted  from  the same  detector
region  during  blank  field  observations were  used  for  background
subtraction.    The  background   subtraction   for  the   high-energy
(non-imaging) instruments  was obtained  by using off-source  data for
the  PDS and  Earth occultation  data  for the  HPGSPC.  Assuming  the
ephemeris as reported  by Stewart et al. (1991)  the first observation
covers the phase interval 0.62--0.84 and the second observation covers
the phase interval 0.11--0.16.

\begin{figure}
\epsscale{0.85}
\plotone{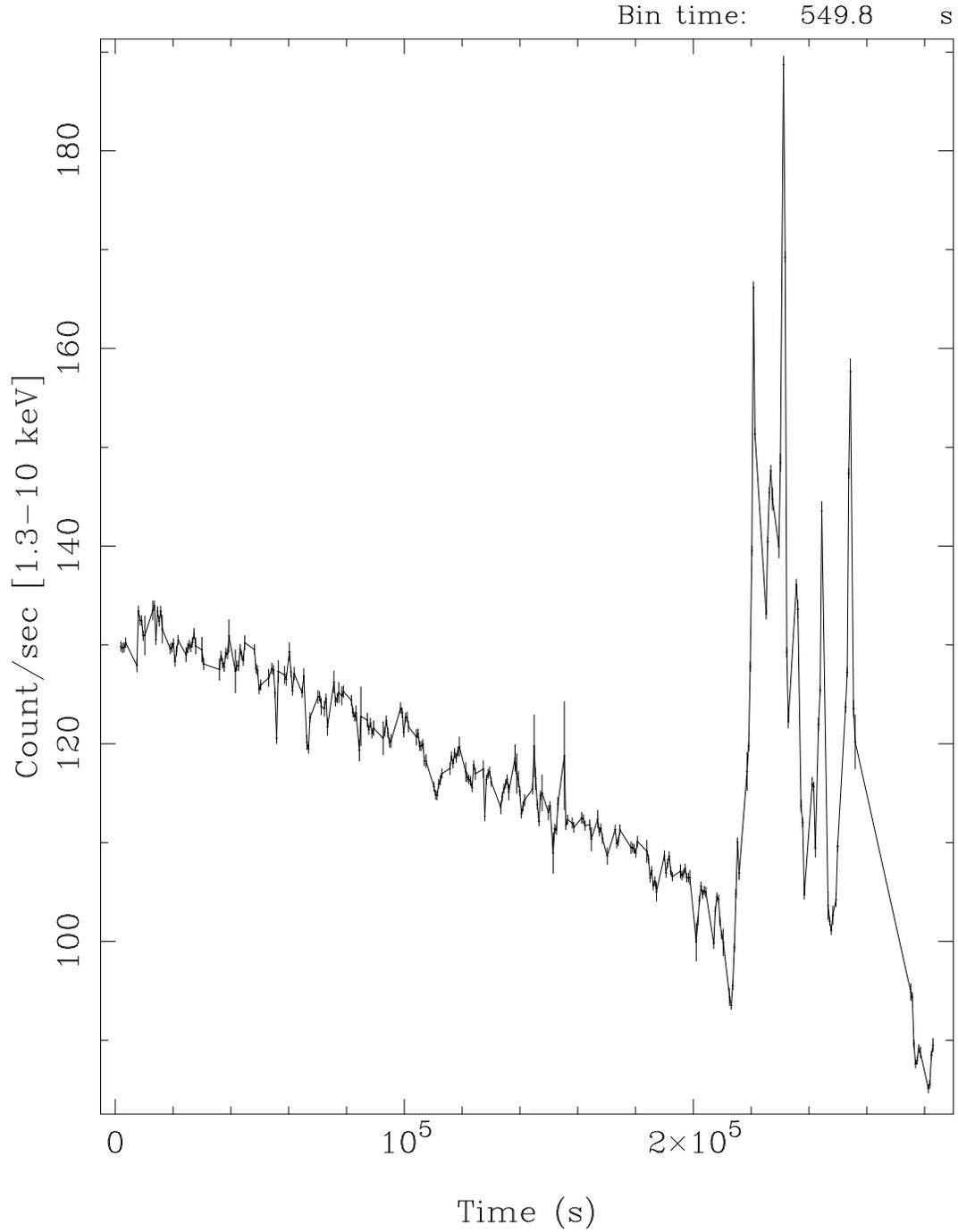}
\caption{  Cir  X--1 lightcurve  in the energy  band 1.3--10 keV
  (MECS data). The data corresponding to the phase interval 0.62--0.84
  assuming the  ephemeris as reported  by Stewart et al.  (1991).  The
  bin time is 550 s. \label{fig1}}
\end{figure}

In Figure~\ref{fig1} we plot the  lightcurve of the observation at the
phase interval 0.62--0.84  in the energy band 1.3--10  keV.  The count
rate  smoothly  decreases from  $\sim  130$  counts  s$^{-1}$, at  the
beginning of the observation, to $\sim 100$ counts s$^{-1}$, at 210 ks
from the beginning of the  observation.  A flaring episode, lasting 50
ks, is  present between  210 ks and  260 ks  from the start  time, the
count  rate varying  from $\sim  100$  counts s$^{-1}$  to $\sim  190$
counts s$^{-1}$.   The lightcurve of  the observation in  the interval
phase  0.11--0.16  shows  a  large  flare  at  the  beginning  of  the
observation lasting around 10 ks and the count rate increases from 300
counts s$^{-1}$ up to $\sim  700$ counts s$^{-1}$; after the flare the
count  rate varies  in a  range between  290 counts  s$^{-1}$  and 400
counts s$^{-1}$ (see Fig. 1 in Iaria et al., 2001a).

\begin{figure}
\plotone{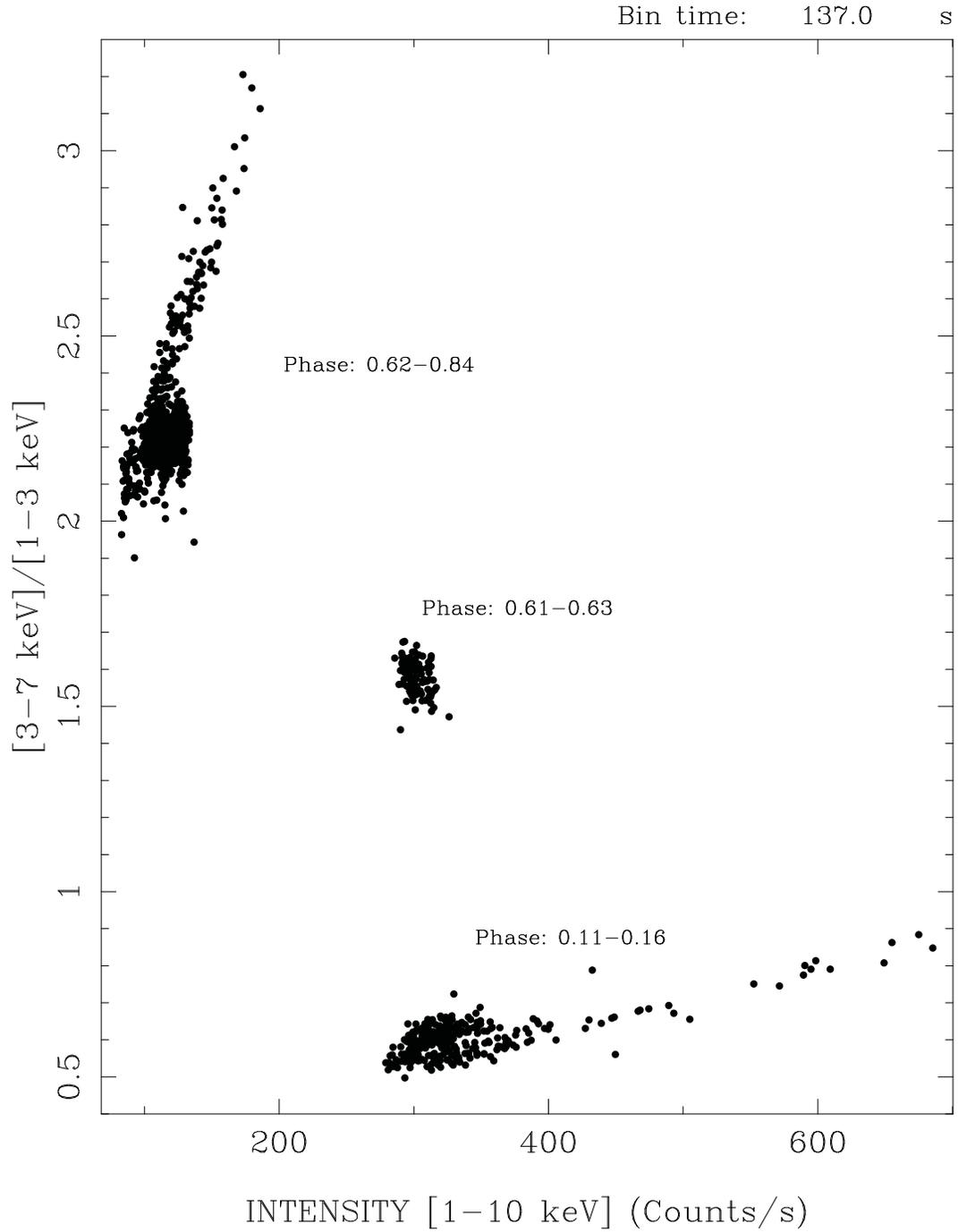}
\caption{ Hardness-intensity diagram of Cir X--1.  We show in
  the same box the data at phase: 0.11-0.16 (Iaria et al. 2001a), the
  data at phase: 0.61-0.63 (Iaria et al. 2002), and the data at phase:
  0.62-0.84 (this paper).  The bin time is 137 s.
\label{fig2}}
\end{figure}

\begin{figure}
\plotone{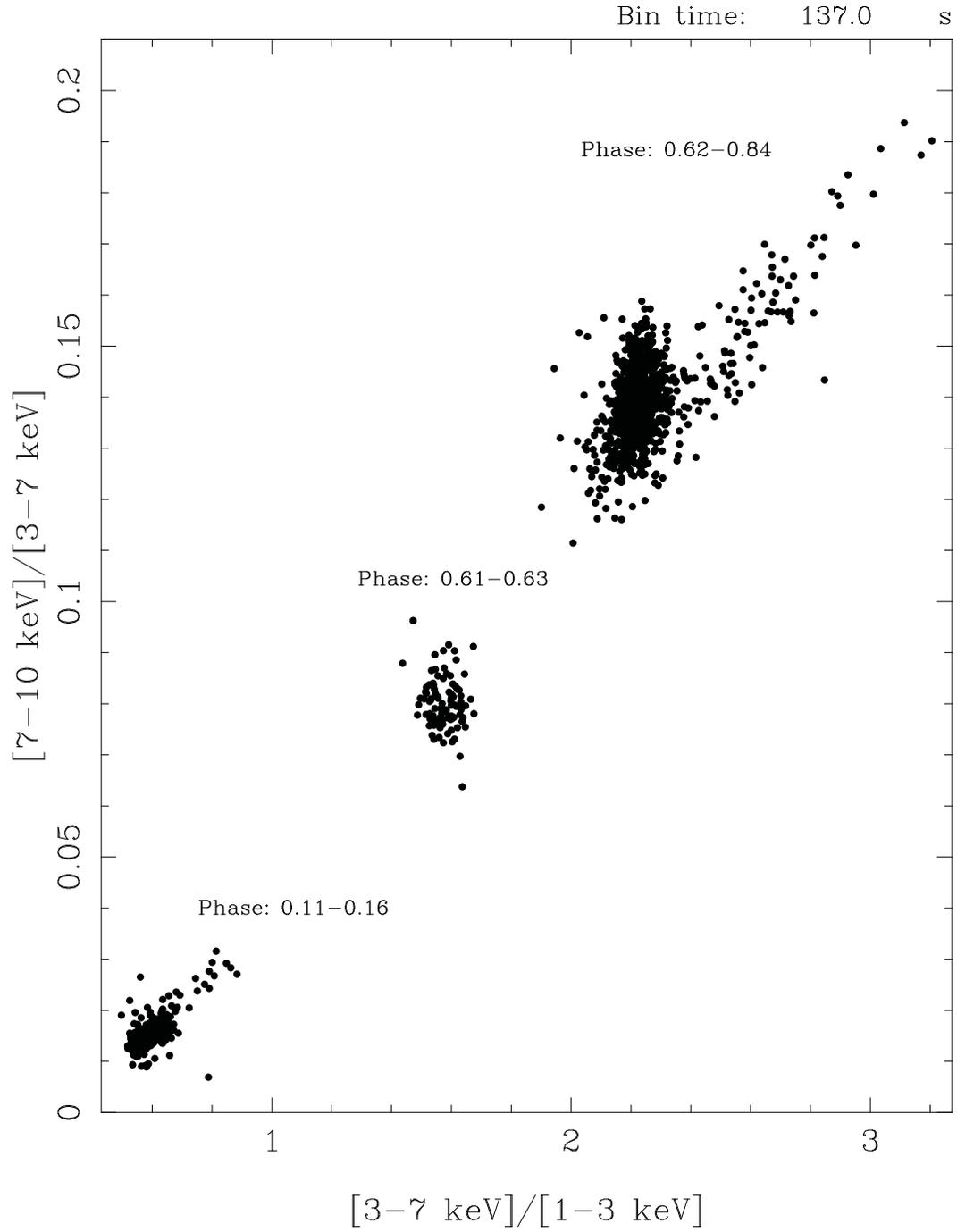}
\caption{ Color-color diagram of Cir X--1 of all the BeppoSAX observations
(same as in Fig.\ref{fig2}).
  The bin time is 137 s.
\label{fig3}}
\end{figure}

In Figure~\ref{fig2} we plot the hardness-intensity diagram, where the
hardness ratio  is the ratio  between the count  rate in the  3--7 keV
energy band  to the count  rate in the  1--3 keV energy band,  and the
intensity is  the count rate  in the 1--10  keV energy band.   In this
figure we show the data of the two observations of this paper together
with  those of a  previous BeppoSAX  observation at  phase 0.61--0.63
(see  Iaria et  al.  2002)  taken in  February 1999.  The data  of the
observation  at  the  interval  phase  0.62--0.84 show  a  lower  mean
intensity of  $\sim 120$ counts  s$^{-1}$ and a larger  hardness ratio
than the  data taken during  the previous observation in  the interval
phase  0.61--0.63, which  shows  a  count rate  of  $\sim 300$  counts
s$^{-1}$  implying  a decreasing  of  the  flux  of $\sim  60$\%.   In
Figure~\ref{fig3} we plot the color-color diagram (CD), where the hard
color (HC)  is the  ratio between  the count rate  in the  energy band
7--10 keV to that in the energy  band 3--7 keV and the soft color (SC)
is the  hardness ratio used  for the hardness-intensity  diagram.  The
data at  phases 0.62--0.84 are at  the top right in  this figure while
the  data of the  previous BeppoSAX  observation at  phases 0.61--0.63
show  lower values  of  both the  hard  and the  soft  color.  The  CD
indicates that the spectrum of Cir X--1, in the energy band 1--10 keV,
is   harder  than   those  corresponding   to  the   previous  BeppoSAX
observations.

To better  see this long-term  variation we checked the  lightcurve of
Cir X--1  taken by the  All Sky Monitor  (ASM) on board of  the X--ray
satellite  RXTE.   During  the   previous  observation  at  the  phase
0.61--0.63  (start time: 51216.14  MJD, stop  time: 51217.80  MJD) the
corresponding  ASM count  rate  is $\sim  86$  counts s$^{-1}$,  while
during the observation analyzed here  it is $\sim 36$ counts s$^{-1}$,
implying  that the  flux  decreased  of 58\%,  in  agreement with  our
BeppoSAX   observations. 

\section{Spectral Analysis}

\begin{figure}
\plotone{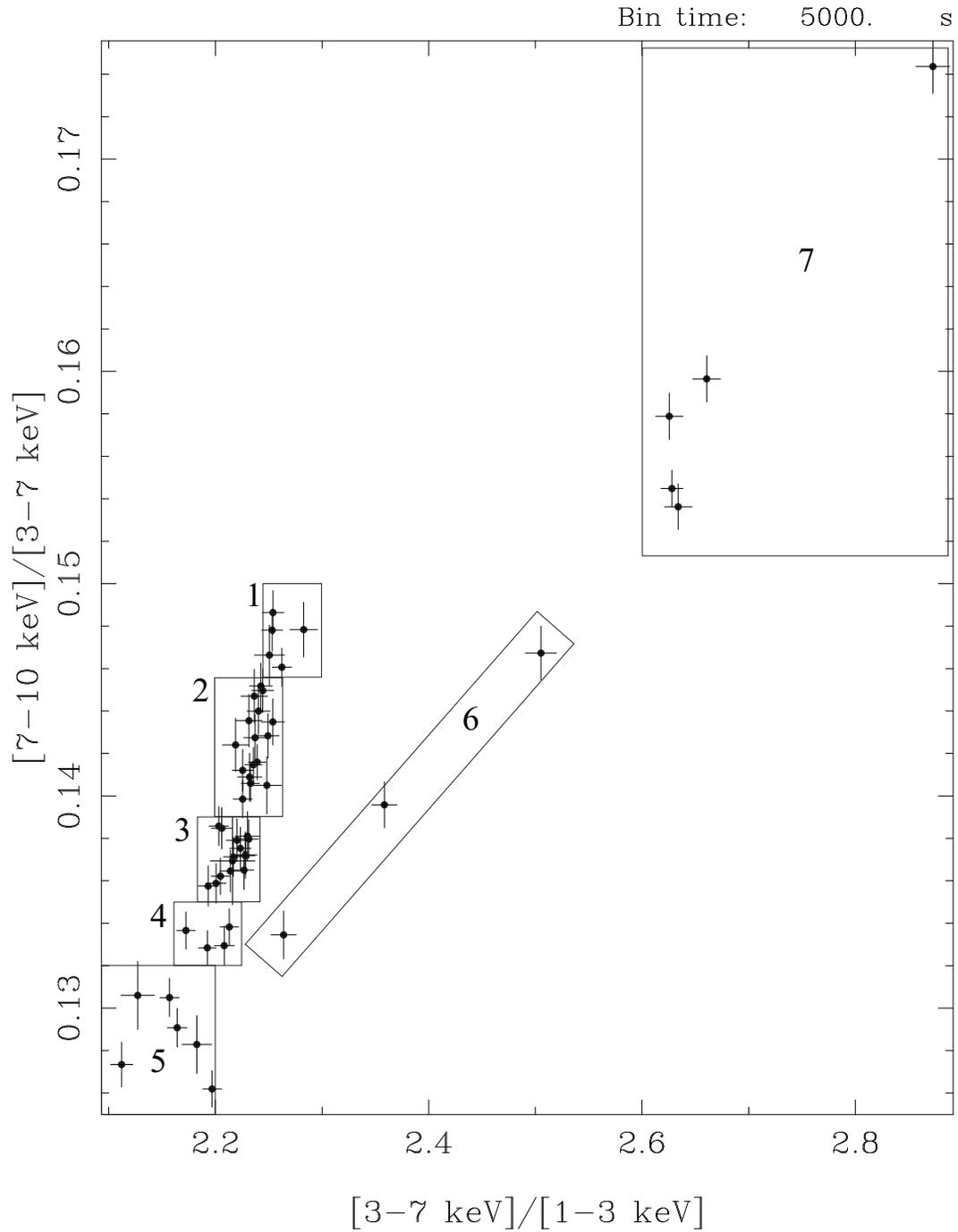}
\caption{ Color-color diagram of Cir X--1 corresponding to the 
  observation at   phases 0.62--0.84.  The boxes indicate the
  seven selected zones from which we extracted the corresponding
  energy spectra. The bin time is 5 ks.
\label{fig5}}
\end{figure}

In Figure~\ref{fig5}  we plot the CD of  the observation corresponding
to the  interval phase 0.62--0.84.   A bin time  of 5 ks was  adopted. 
The seven zones  used to extract the corresponding  energy spectra are
indicated  in the  figure; zones  6 and  7 correspond  to  the flaring
episode. In Tab. \ref{tab1} we report the intervals of SC and HC which
identify the seven selected zones,  and the exposure times for each of
the  instruments.   For  each of  the  selected  zones  in the  CD  we
extracted energy  spectra for each instrument, which  were rebinned in
order to have at least  25 counts/energy-channel and to oversample the
instrumental energy resolution with the same number of channels at all
energies\footnote{see       the       BeppoSAX       cookbook       at
  http://www.sdc.asi.it/software/index.html}.   A systematic  error of
1\%  was added  to the  data. As  customary, in  the  spectral fitting
procedure we allowed for a different normalization of the LECS, HPGSPC
and PDS  spectra relative to  the MECS spectrum, always  checking that
the derived values are in  the standard range for each instrument. The
energy ranges used  for the spectral analysis are  0.12--4 keV for the
LECS, 1.8--10 keV  for the MECS, 7--30 for the  HPGSPC and 15--200 keV
for the PDS.   We indicate the seven spectra with numbers  from 1 to 7
corresponding to the seven selected zones in the CD.
 
\begin{figure}
\plotone{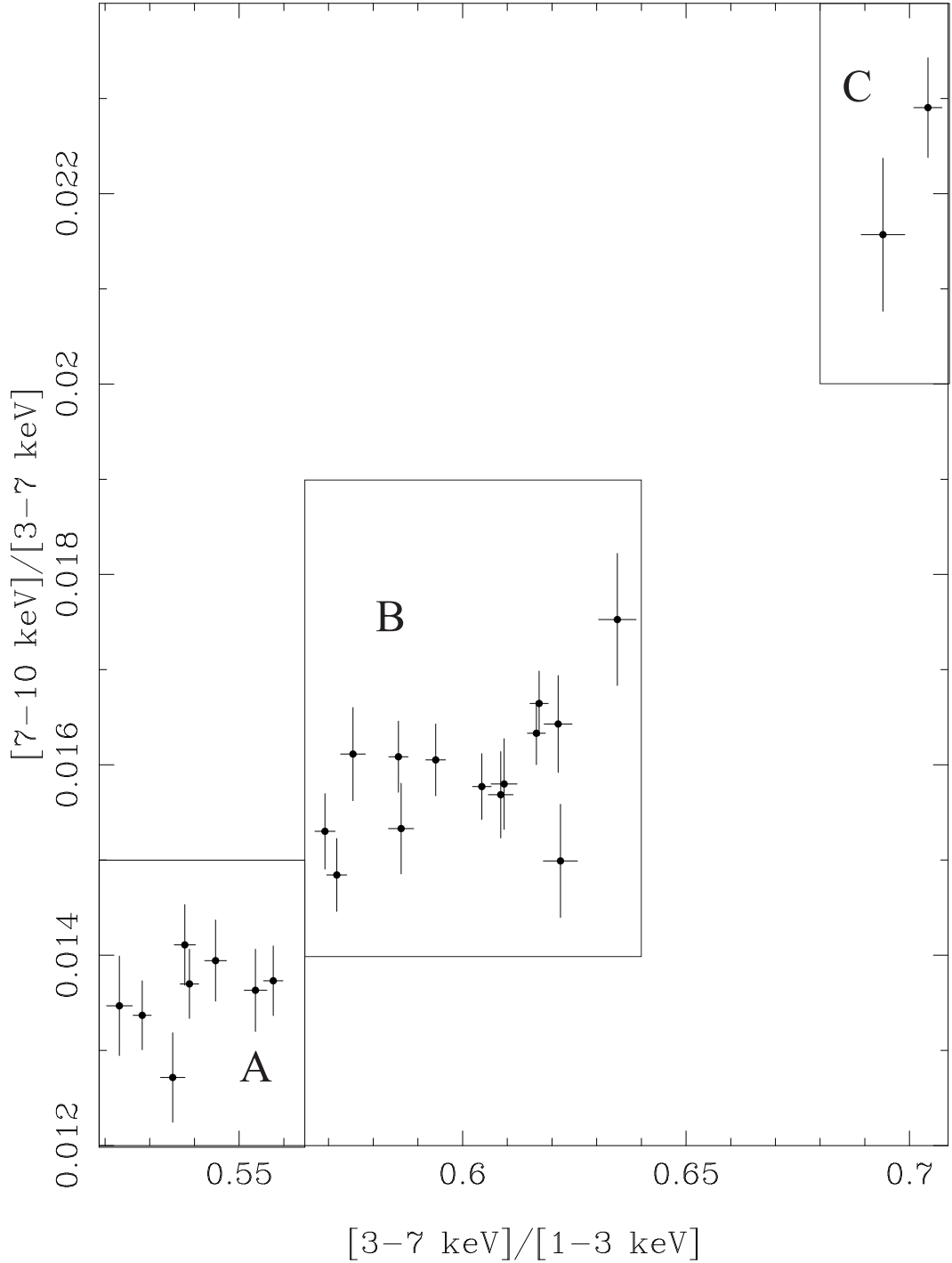}
\caption{ Color-color diagram of Cir X--1 corresponding to the 
  observation at   phases 0.11--0.16.  The  boxes indicate
  the selected regions  from which we extracted the corresponding
  energy spectra.  The bin time is 3 ks.
\label{fig2per}}
\end{figure}

In Figure~\ref{fig2per} we plotted the CD of the observation at phases
0.11--0.16.   This observation  was already  studied by  Iaria et  al. 
(2001) which extracted from eight temporal intervals the corresponding
energy  spectra,  grouping the  data  in order  to  have  at least  20
counts/energy  channel.   In this  paper  we  apply  to the  data  the
logarithmic  grouping  (in order  to  oversample  to the  instrumental
energy resolution  with the same  number of channels at  all energies)
and a systematic error of 1\%, instead of 2\% as done in the previous
analysis, (see e.g.   Frontera et al., 2001).  From the CD we selected
three zones  (see Fig.~\ref{fig2per}); the  zone C corresponds  to the
flaring episode. In Tab. \ref{tab2}  we report the intervals of SC and
HC which identify the three selected zones, and the exposure times for
each  of  the  instruments.   We  extracted energy  spectra  for  each
instrument  from zone  A  and B  because  the zone  C  has not  enough
statistics.  The spectra were  extracted using the procedure described
above.   The  energy  ranges   used  for  the  spectral  analysis  are
0.12--2.85 keV for  the LECS, 1.8--10 keV for the  MECS, 7--30 for the
HPGSPC and  15--200 keV for the  PDS.  We indicate  the energy spectra
with  letters A  and B  corresponding, respectively,  to  the selected
zones A and B in the CD.
 

\subsection{The soft excess  below 1 keV}

\label{baseapo}
\begin{figure}
\centering
 \includegraphics[width=10cm]{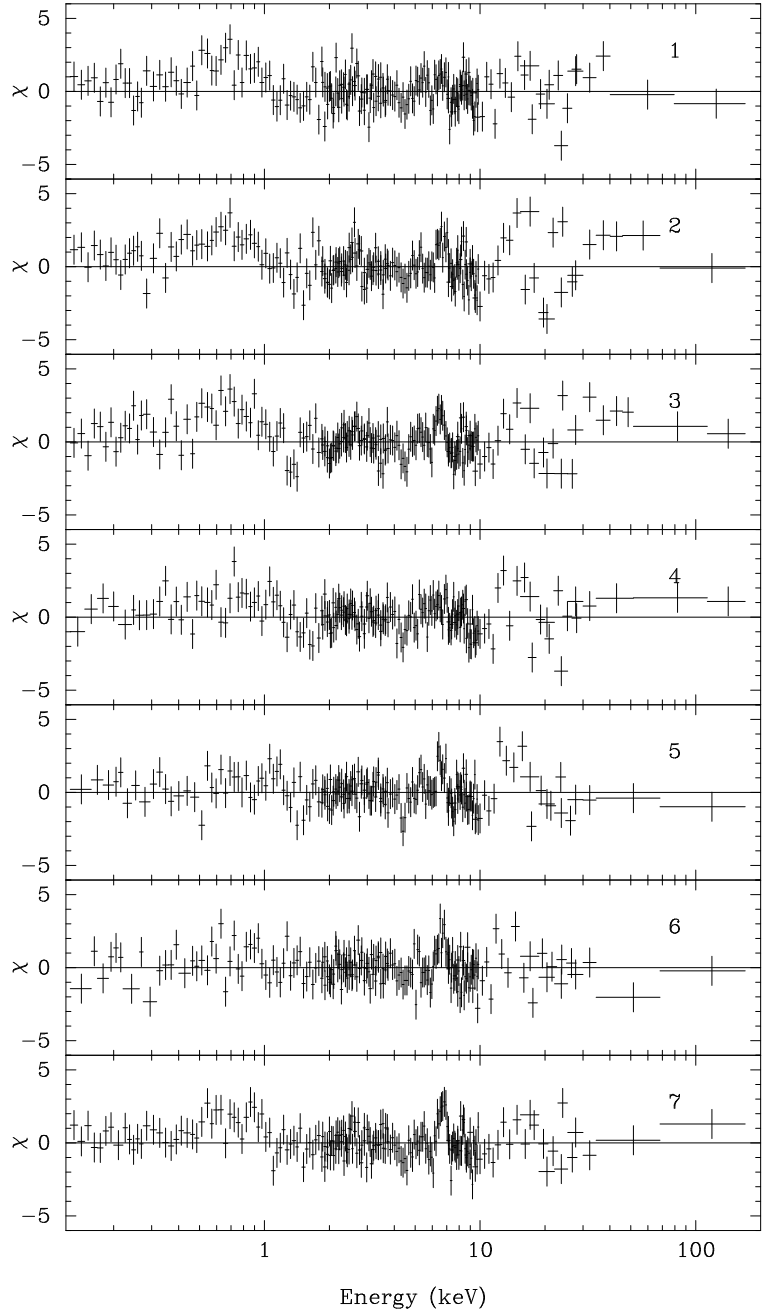}
\caption{ Residuals, in units of $\sigma$, of the seven  spectra corresponding
  to the observation at the   phases 0.62--0.84.  The
  residuals are with respect to the model reported in  Tab. \ref{Tabbaseapo}.
\label{fig6}}
\end{figure}

To fit  the seven  spectra corresponding to  the observation  taken at
phases  0.62--0.84  we  used  the Comptonization  model  {\it  Comptt}
(Titarchuk,  1994) to  which we  added  a blackbody  component at  low
energies in agreement with Iaria et al.  (2002) who studied the source
at  similar phases.   This model  gave a  $\chi^2/d.o.f.$  of 285/202,
364/200,   330/200,   254/201,    244/199,   249/199,   and   256/200,
respectively, for  spectra 1 to  7; in Table \ref{Tabbaseapo}  we show
the  corresponding  best fit  parameters.   The equivalent  absorption
hydrogen column was $\sim 1.7 \times 10^{22}$ cm$^{-2}$, the blackbody
temperature  around  0.55 keV,  the  seed-photon  temperature and  the
electron temperature of the Comptonized component around 1 keV and 2.7
keV,  respectively, and  finally  the optical  depth,  $\tau$, of  the
Comptonizing  cloud  is 11.   The  residuals  obtained  for the  seven
spectra are  shown in Figure~\ref{fig6}.   The residuals corresponding
to spectra  1, 2, 3  and 7 show  a prominent soft excess  reaching the
maximum at 0.6--0.7  keV.  The excess is less evident  in spectra 4, 5
and 6, probably because of the  lower statistics due to the lower LECS
exposure times in these spectra  (see Tab. 1).

\begin{figure}
\plotone{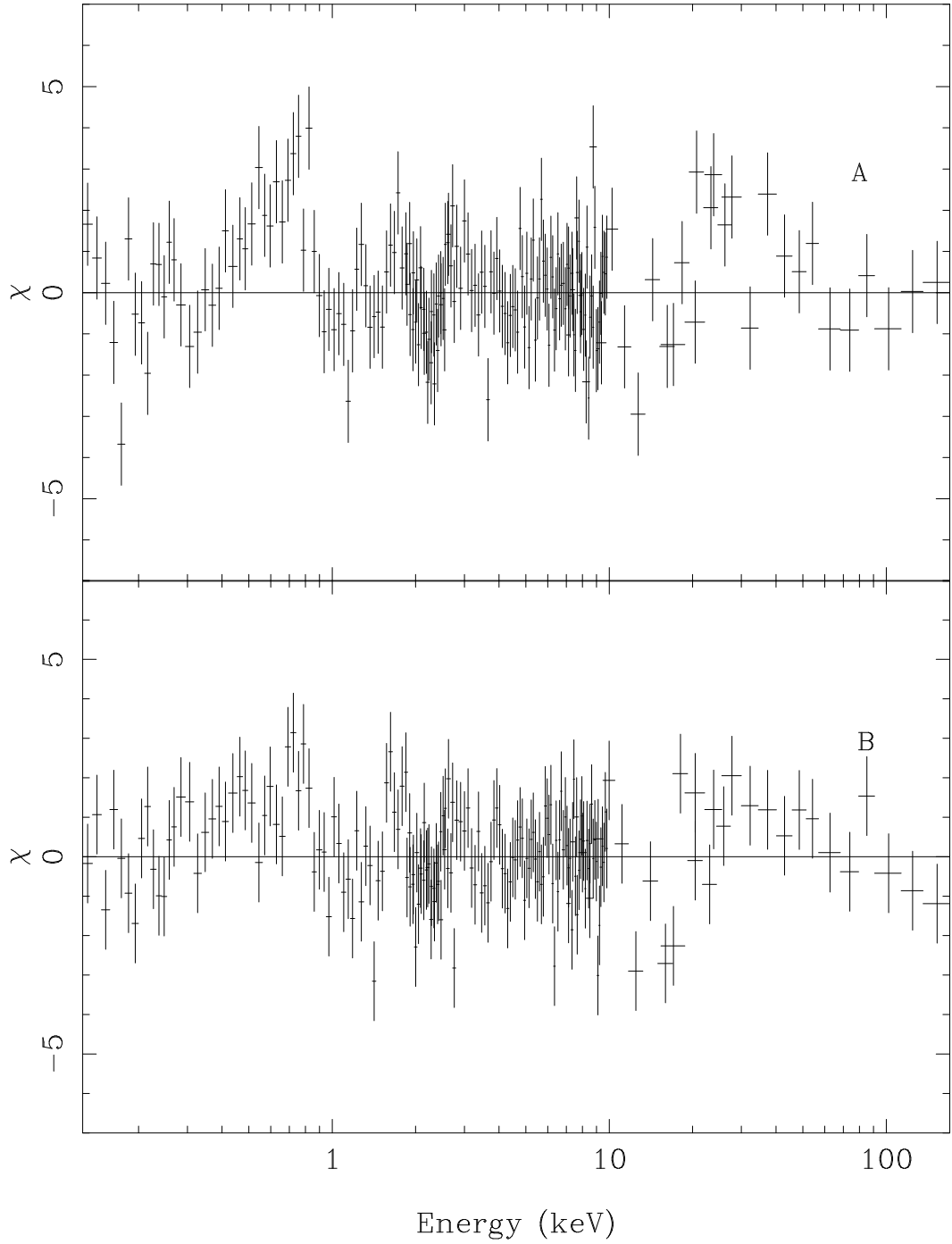}
\caption{ Residuals, in units of $\sigma$, of the  spectra A e B corresponding
  to the observation at the phases 0.11--0.16 with respect to the
  model adopted by Iaria et al. (2001a).
\label{figresbadperi}}
\end{figure}

To  fit the  spectra A  and B  (phases 0.11--0.16)  we used  the model
proposed by Iaria et al.  (2001a).  The fits gave a $\chi^2/d.o.f.$ of
316/176 and 260/175, respectively for  spectrum A and B. The residuals
corresponding     to    the    best     fits    are     reported    in
Fig.~\ref{figresbadperi}.  We  note again the presence  of a prominent
soft excess in the residuals  of both the spectra reaching the maximum
at  0.6--0.7  keV  similarly   to  those  obtained  from  the  spectra
corresponding  to the  observation taken  at phases  0.62--0.84.  This
soft excess is less prominent in spectrum B, again probably due to the
lower statistics  due to  the lower LECS  exposure time in  spectrum B
(see Tab. \ref{tab2}).

\subsection{Fitting of the soft excess below 1 keV}

To  study the  soft excess  below  1 keV  we start  fitting the  seven
spectra corresponding  to the  observation taken at  phases 0.62--0.84
because  of   the  higher   statistics  (see  Tabs.    \ref{tab1}  and
\ref{tab2}). Then  we discuss several possible models,  and finally we
apply the best model to spectra A and B taken at phases 0.11--0.16.
 
Previous ASCA spectra of Cir X--1 were fitted using a partial covering
component (Brandt et al., 1996;  Iaria et al., 2001b).  To compare our
results with the previous ASCA results we fitted the seven spectra, in
the energy  range 1--10 keV, using  only LECS and MECS  data, with the
model  reported  by  Brandt  et   al.   (1996).   This  model  gave  a
$\chi^2/d.o.f.$ of 113/124,  119/124, 96/124, 95/124, 94/124, 104/124,
and  103/124, respectively,  for spectra  1 to  7.  In  agreement with
Brandt  et al.  (1996)  we find  that the  continuum emission  is well
fitted by two blackbody  components absorbed by an equivalent hydrogen
column of  $\sim 1.7 \times 10^{22}$ cm$^{-2}$.   The partial covering
component  is present in  the seven  spectra, the  equivalent hydrogen
column,  $N_{\rm H_{pc}}$,  associated to  this component  is  $\sim 2
\times  10^{23}$ cm$^{-2}$ and  the covered  fraction of  the emitting
region is $\sim  0.3$. We applied the same model  to our spectra using
the energy  range 0.12--10  keV taking into  account the  energy range
where the  soft excess is  present.  This model  gave bad fits  with a
$\chi^2/d.o.f.$  of  252/164,   308/164,  373/164,  177/164,  198/164,
198/164, and 204/164,  respectively, for spectra from 1  to 7, lefting
unchanged the  residuals below 1 keV.   Finally we fitted  the data in
the  whole energy  range 0.12--200  keV using  the  continuum emission
described  in  section~\ref{baseapo} to  which  we  added the  partial
covering component.   We found that the  spectrum above 1  keV is well
fitted but  the soft  excess below 1  keV persists, implying  that the
soft  excess cannot  be fitted  by  a partial  covering model.   These
results clearly  show that  the spectrum of  Cir X--1 is  more complex
than it appeared in the energy band 1--10 keV.

We tried  several emission components to  fit this soft  excess but no
one  gave reasonable  results.   The addition  of  an other  blackbody
component with a temperature of  $\sim 0.15$ keV improved the fit, but
the corresponding  unabsorbed luminosity  was $\sim 10^{42}$  erg/s in
the 0.1--200  keV energy range.  This  result is very  hard to explain
assuming that  the X--ray binary system  contains a neutron  star or a
Galactic black hole.  Instead of  a blackbody we added a Gaussian line
centered  at  0.6  keV.   Also  in  this case  the  results  were  not
physically   acceptable  because,   due   to  the   high  N$_H$,   the
corresponding equivalent width of the line was larger than 1 keV.

We conclude  that the  soft excess may  have two possible  origins: a)
there  is a continuum  emission component  below 1  keV absorbed  by a
lower value  of equivalent hydrogen  column with respect to  the N$_H$
absorbing the spectrum above 1  keV; b) the equivalent hydrogen column
of the whole spectrum is overestimated.

As regards the  first hypothesis, the residuals below  1 keV disappear
if  we  add to  the  previous  model  a continuum  emission  component
absorbed by a different lower  equivalent hydrogen column (N$_H \sim 8
\times 10^{20}$ cm$^{-2}$) than that absorbing the other components of
the model ($\sim 1.7 \times  10^{22}$ cm$^{-2}$). However in this case
it is  hard to  explain the  lower value of  N$_H$ since  the emitting
source should  be distant less  than 1 kpc.   Because in the  LECS and
MECS  field of  view  no  contaminating source  is  visible we  should
conclude that  Cir X--1 is  distant less 1  kpc or that a  source with
coordinates compatible with those of Cir X--1, at a distance less then 1
kpc, is present.  Since this scenario looks unrealistic, in this paper
we concentrate on the second hypothesis.

The  second  hypothesis would  imply  an  equivalent hydrogen  column,
N$_H$, lower than $\sim 1.7  \times 10^{22}$ cm$^{-2}$.  A good fit of
the broad band spectra is  obtained adding two absorption edges at 0.7
keV and 1--1.2 keV having a large optical depth of $\sim 5 $ and $\sim
1 $ respectively.   The threshold energy of the  first absorption edge
is similar in all the seven  spectra while the threshold energy of the
second absorption edge varies between 1 keV and 1.2 keV.  Adding these
features we find  that the equivalent hydrogen column  N$_H$ is $ \sim
0.6 \times  10^{22}$ cm$^{-2}$  and the spectrum  below 1 keV  is well
fitted.

\begin{figure}
\centering
 \includegraphics[width=10cm]{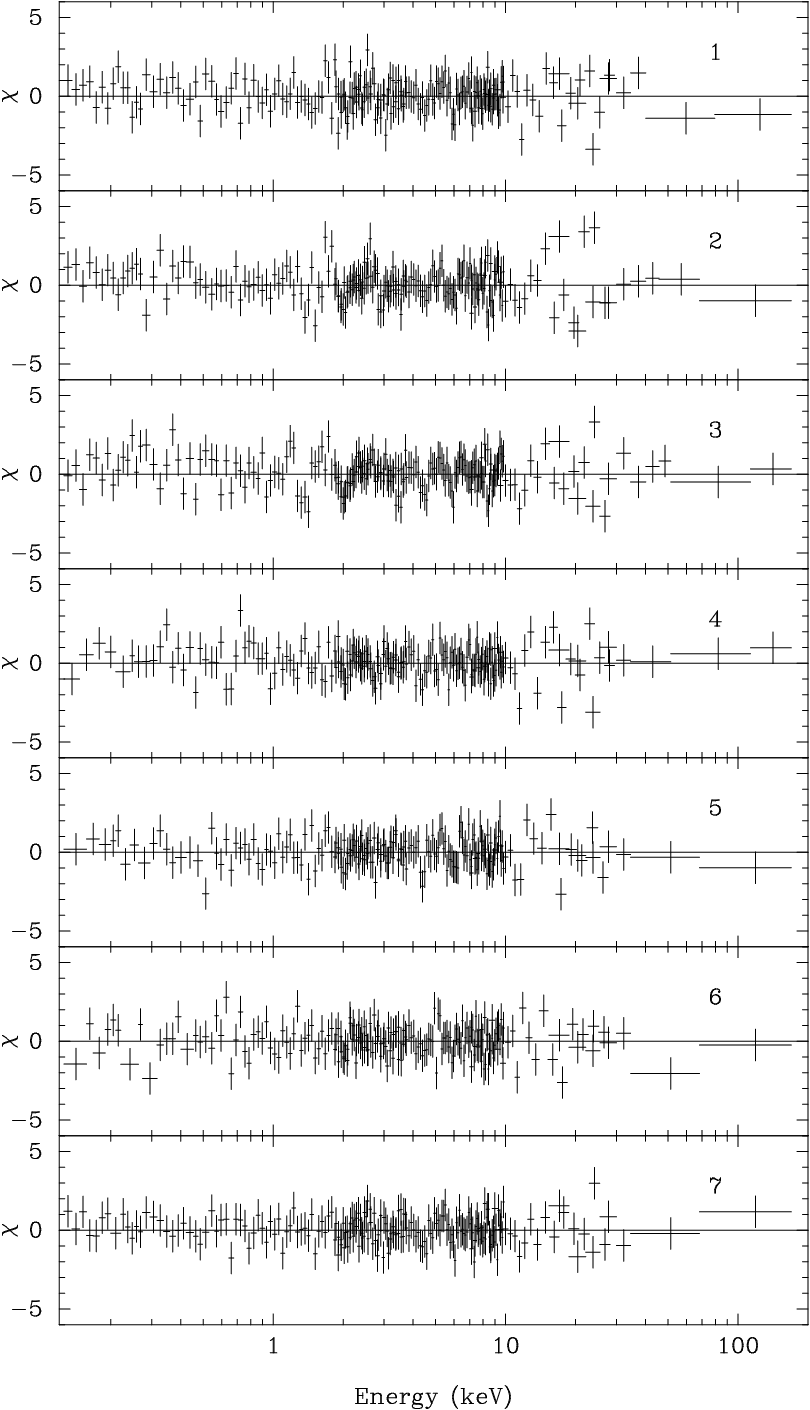}
\caption{Residuals, in units of $\sigma$, of the seven  spectra corresponding
  to the observation at the   phases 0.62--0.84.  The residuals
  are with respect to the model reported in Tab. \ref{tabnifeapo}.
  \label{fig9}}
\end{figure}

In  the residuals of  spectra 2  and 3  a hard  excess is  present. In
agreement  with previous  works (Iaria  et al.,  2001a; Iaria  et al.,
2002; Ding  et al., 2003), we  fitted it adding  a power-law component
having  a photon  index fixed  to 2.6  in all  the seven  spectra.  In
spectra 6 and  7 a gaussian emission line, having  its centroid at 6.7
keV, was  needed.  Two localized features  are present at 7  keV and 9
keV in the  seven spectra, we added two absorption edges  to fit them. 
The  residuals  with  respect  to  the final  model  are  reported  in
Figure~\ref{fig9},  the  best fit  parameters  are  reported in  Table
\ref{tabnifeapo}.

Because the structure of the residuals  below 1 keV is the same in the
spectra  of both the  observations presented  in this  work we  fitted 
spectra A and B, corresponding  to the observation taken at the phases
0.11--0.16, adding  two absorption  edges at 0.6  and 0.8 keV,  in the
model  adopted by  Iaria et  al. (2001a).  In this  case the  fits are
improved adding two further absorption edges at 1.3 and 1.9 keV.

\begin{figure}
\plotone{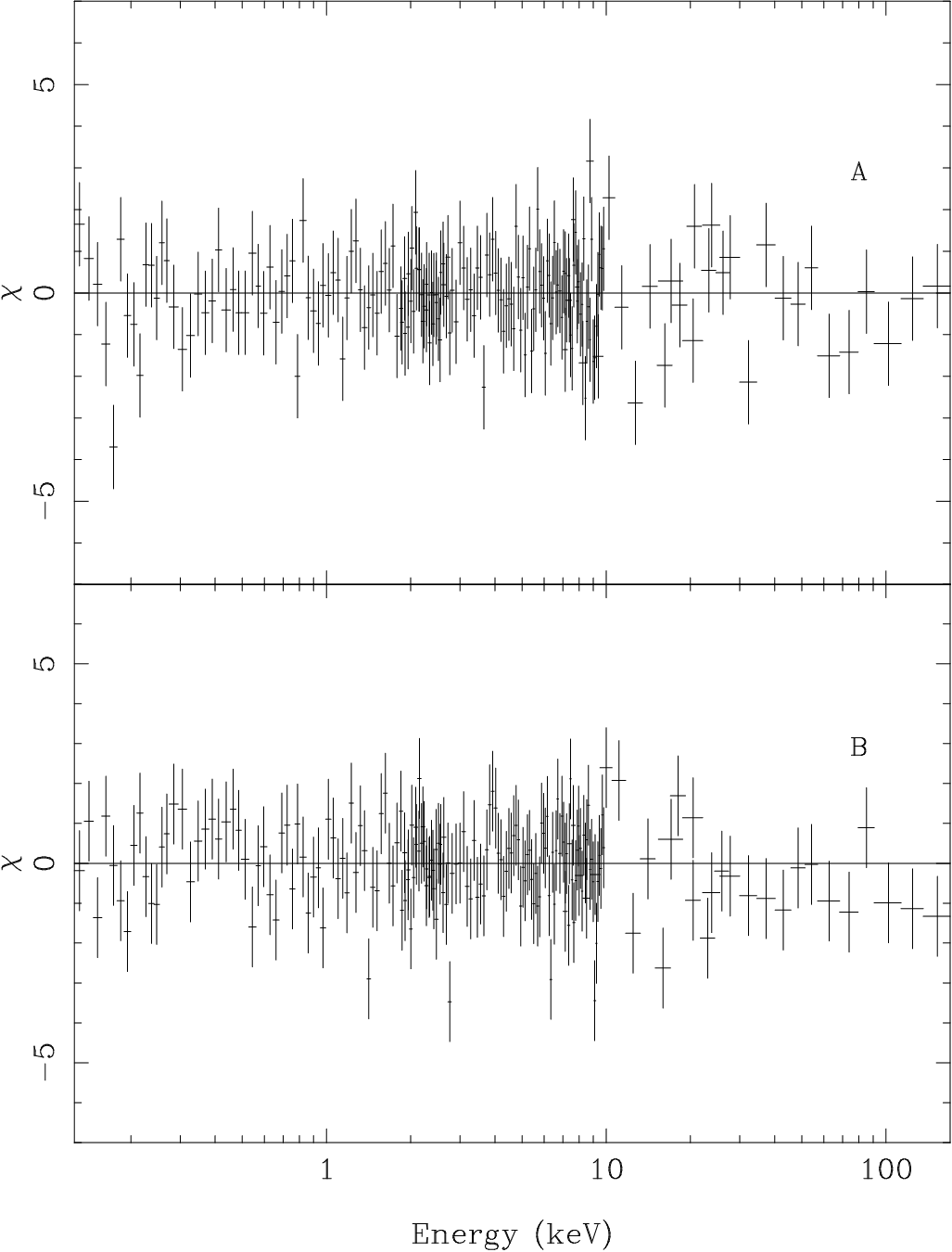}
\caption{ Residuals, in units of $\sigma$, of the  spectra A e B
  corresponding  to the  observation  at the  phases 0.11--0.16.   The
  residuals  are  with   respect  to  the  model  reported   in  Tab.  
  \ref{tabnifeperi}.
\label{fignifeperi}}
\end{figure}

The parameters  of these fits  are reported in  Tab. \ref{tabnifeperi}
and the residuals  corresponding to the best-fits are  plotted in Fig. 
\ref{fignifeperi}.   The addition  of four  absorption edges  does not
significantly  change the parameters  associated to  the Comptonized
component  and  the  power-law  component.  The photon  index  of  the
power-law is fixed to 2.6 (as in spectra from 1 to 7).  The equivalent
hydrogen column  is $ \sim  0.6 \times 10^{22}$  cm$^{-2}$, compatible
with the values obtained from the other observation.

\begin{figure}
\centering
 \includegraphics[width=10cm]{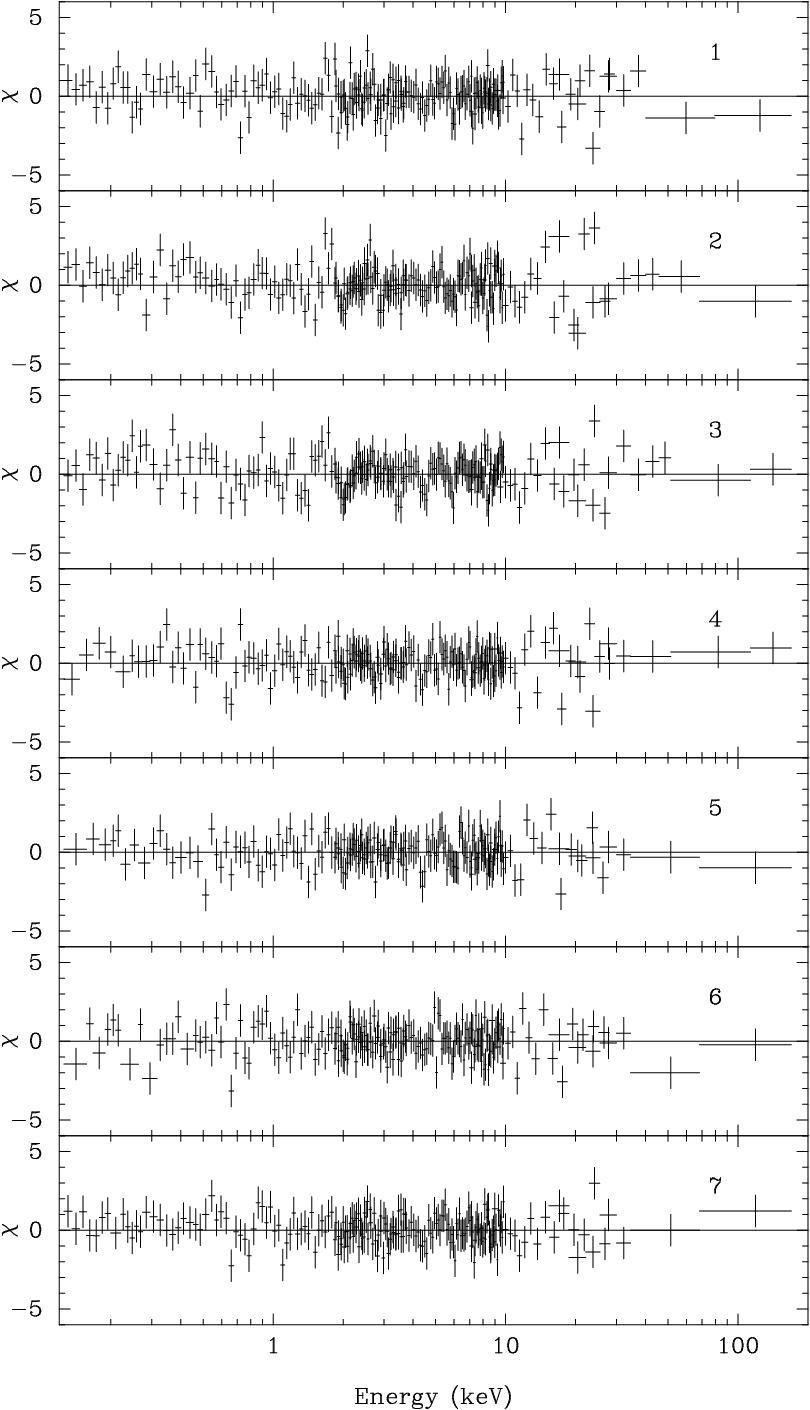}
\caption{Residuals, in units of $\sigma$, of the seven  spectra corresponding
  to the observation at the   phases 0.62--0.84.  The residuals
  are with respect to the model reported in Tab. \ref{tabapofix}.
  \label{figapofix}}
\end{figure}

\begin{figure}
\plotone{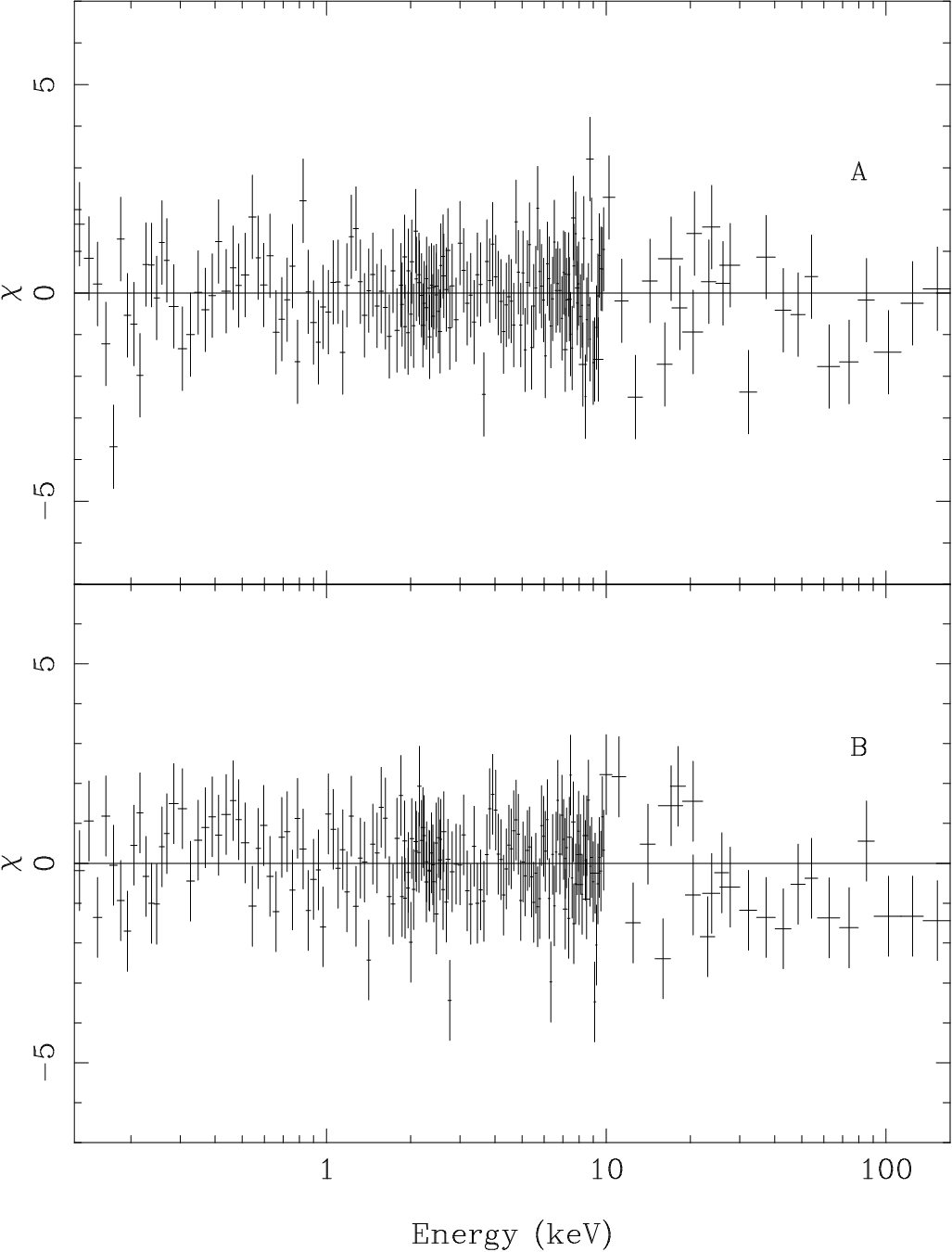}
\caption{Residuals, in units of $\sigma$, of the two  spectra corresponding
  to the observation at the   phases 0.11--0.16.  The residuals
  are with respect to the model reported in Tab. \ref{tabperifix}.
  \label{figperifix}}
\end{figure}

The poor LECS energy resolution and  the low statistics below 1 keV do
not allow us to well identify the absorption features at low energies.
For this reason  we fit the spectra from 1 to  7 fixing the equivalent
hydrogen column  to $0.66 \times 10^{22}$ cm$^{-2}$  and the threshold
energies of  the absorption edges  to those of  \ion{O}{7}, \ion{O}{8}
and \ion{Ne}{9}.   We use  the same continuum  model reported in  Tab. 
\ref{tabnifeapo}.   In Tab.   \ref{tabapofix} we  report  the best-fit
parameters  and in  Fig. \ref{figapofix}  we show  the  residuals with
respect to this  model.  We note that the  parameters of the continuum
components are  unchanged, and  that we obtain  good fits for  all the
seven spectra.   Finally we fit  the spectra A  and B also  fixing the
equivalent hydrogen column to  $0.66 \times 10^{22}$ cm$^{-2}$ and the
threshold energies  of the absorption edges  associated to \ion{O}{7},
\ion{O}{8}, \ion{Mg}{11}  and \ion{Mg}{12}. In this  case two gaussian
absorption lines  associated to \ion{O}{8} and  \ion{Mg}{12} are added
at $\sim  0.65$ keV  and $\sim 1.47$  keV, respectively. Also  in this
case we find  good fits, the best-fit parameters  of the continuum are
unchanged   with   respect   to   the   model   reported   in   Tab.   
\ref{tabnifeperi};   the   new   results   are  reported   in   Tab.   
\ref{tabperifix}   and   the   residuals   are   plotted   in   Fig.   
\ref{figperifix}.

\section{Discussion}
We analyzed data  of Cir X--1 from two  BeppoSAX observations taken in
September  2001 and August  1998 at  phases 0.62--0.84  and 0.11--0.16
respectively.

About  the  observation  at  phases  0.62--0.84, taken  in  2001,  the
lightcurve, the hardness-intensity diagram and the color-color diagram
show that  the intensity has decreased  and soft and  hard colors have
increased  with respect  to  previous observations  at similar  phases
(0.61--0.63,  Iaria et  al. 2002).   We  selected seven  zones in  the
color-color diagram and extracted  from each of them the corresponding
spectrum.  The spectra,  in the energy band 1--10  keV, are similar to
that of  other Z-sources (see Di  Salvo et al.  2000; Di Salvo et  al. 
2001, Iaria  et al. 2003),  and are well  fitted by a blackbody  and a
Comptonized component.  Below 1  keV a soft  excess is present  in the
residuals of spectra 1, 2, 3  and 7 (see Fig. \ref{fig6}); this can be
fitted adding two  absorption edges at 0.7 and  1 keV.  The equivalent
hydrogen column  density, N$_H$,  absorbing the continuum  emission is
$\sim 0.6 \times 10^{22}$ cm$^{-2}$.  In spectra 4, 5 and 6, where the
soft excess  is not evident, we  fixed N$_H$ to $  0.6 \times 10^{22}$
cm$^{-2}$  and added two  absorption edges  at low  energies obtaining
similar results.   For this reason  we conclude that in  these spectra
the  absence of  the  soft excess  below 1  keV  is due  to the  lower
statistics in the  LECS data (see Tab.  \ref{tab1}).  However the poor
LECS energy resolution and the low statistics below 1 keV do not allow
us to well identify the absorption features at low energies.  For this
reason  we  fitted the  spectra  from 1  to  7  fixing the  equivalent
hydrogen column  to $0.66 \times 10^{22}$ cm$^{-2}$  and the threshold
energies of  the absorption edges  to those of  \ion{O}{7}, \ion{O}{8}
and \ion{Ne}{9}. In the following we will discuss this latter model.

About  the  observation  at   phases  0.11--0.16  taken  in  1998,  we
reanalysed data already published  by Iaria et al.  (2001a) extracting
the spectra  from the corresponding  color-color diagram while  in the
previous  paper  the  spectra   were  extracted  from  eight  temporal
intervals  along  the  entire  observation.   This allows  to  put  in
evidence the soft  excess below 1 keV, only  marginally visible in the
previous analysis (see  Fig. 3 in Iaria et  al.  2001a). The continuum
used to fit the data and the corresponding best-fit parameters are the
same  as reported  by Iaria  et al.   (2001a).  The  addition  of four
absorption edges at 0.6 keV, 0.8  keV, 1.3 keV and 1.9 keV changes the
equivalent absorption column that it  is now $\sim 0.6 \times 10^{22}$
cm$^{-2}$ instead  of $\sim 1.7 \times 10^{22}$  cm$^{-2}$ as reported
in the previous  paper. Also in this analysis,  a prominent absorption
edge at 8.4 keV and a Gaussian  line at $\sim 6.7 $ keV are present as
already discussed by  Iaria et al. (2001a). Also  for this observation
the poor LECS energy resolution and  the low statistics below 1 keV do
not allow us to well identify the absorption features at low energies.
For this  reason we fitted the  spectra A and B  fixing the equivalent
hydrogen column  to $0.66 \times 10^{22}$ cm$^{-2}$  and the threshold
energies of the absorption edges associated to \ion{O}{7}, \ion{O}{8},
\ion{Mg}{11} and  \ion{Mg}{12}.  In this case  two gaussian absorption
lines  associated to \ion{O}{8}  and \ion{Mg}{12}  are added  at $\sim
0.65$ keV and $\sim 1.47$  keV, respectively. In the following we will
discuss this latter model.

This modeling  was possible thanks  to the good spectral  coverage of
the BeppoSAX/LECS, which  gives us the possibility to  study the X-ray
spectrum  down to 0.1  keV.  For  instance, recent  Chandra observations
(Brandt \&  Shulz 2000; Shulz  \& Brandt 2002)  did not allow  to well
determine  the  continuum and  the  absorption  features  below 1  keV
because the useful spectral range  was limited to 1.4--7.3 keV. On the
other hand, a  K-shell absorption edge of neutral  iron was observed in
the two Chandra observations, although  the choice to fix the hydrogen
absorption  column   to  $2  \times  10^{22}$   cm$^{-2}$  might  have
influenced the  value of the optical  depth of the  edge obtained from
the  fit.  In a  previous ASCA  observation (Brandt  et al.  1996) the
estimated equivalent hydrogen column was N$_H \sim 1.7 \times 10^{22}$
cm$^{-2}$  similar  to the  value  that  we  found not  including  the
absorption edges  at low  energies (see Tab.  \ref{Tabbaseapo}).  Note
that  the ASCA energy  band used by  Brandt et al. (1996)  is 1--10
keV, which  does not  allow a clear  detection of  possible absorption
edges at 0.7--1  keV. We confirm their results in the   energy
range 1--10 keV, where the spectrum  is well fitted with an
N$_H  \sim  1.7  \times  10^{22}$  cm$^{-2}$  and  where  the  K-shell
absorption edge  associated to neutral  iron is fitted using  a partial
covering  component. However, extending the energy range down to 0.1 keV
this model does not fit the data below 1 keV.

\subsection{The distance to Cir X--1}
A source  having the position  of Cir X--1  and a distance to  5.5 kpc
(Case \&  Battacharya 1998) has  a corresponding visual  extinction of
$5.8  \pm 2.0$  mag  (Hakkila et  al.  1997) and,  using the  observed
correlation between  visual extinction and  equivalent hydrogen column
(Predehl \& Schmitt 1995), an  equivalent hydrogen column N$_H = (1.03
\pm 0.02) \times  10^{22}$ cm$^{-2}$. In the previous  ASCA (Brandt et
al. 1996; Iaria et al. 2001b)  and BeppoSAX (Iaria et al. 2001a; Iaria
et al.  2002) observations an  equivalent absorption column of  $ \sim
1.7 \times 10^{22}$ cm$^{-2}$ was found and the authors concluded that
an excess  of obscuring matter was present  close to the X-ray
source.  The value  of the N$_H$ obtained in  this analysis is smaller
than the  previous one by  a factor of  three. The absorption  we find
(Tabs. \ref{tabnifeapo} and \ref{tabnifeperi})
implies  that  for  a source  having  the  position  of Cir  X--1  the
corresponding distance should be around 4 kpc.

If  the interpretation of  our data  is correct  the debate  about the
distance  of  Cir  X--1  now  becomes  complicated.   The  N$_H$-based
distance is not compatible with a distance between 6.4--8~kpc obtained
from the 21-cm absorption features in the radio spectrum of the source
(Goss \& Mebold  1977; Glass 1994).  Now we  discuss this result using
the latest  estimate of  the distance $R_0  = 8.5  \pm 0.5$ kpc  of the
Galactic center  from the Sun (Feast \&Whitelock  1997) obtained using
Hipparcos data.
 
Looking at the absorption \ion{H}{1}  spectrum in the direction of Cir
X--1  (see figure  1 in  Goss \&  Mebold 1977)  five  prominent radial
velocities are present at 3.0, -6.5,  -21.0, -57.5 and -75 km/s with a
resolution of 0.82 km/s, the latter  feature extends down to -90 km/s. 
Goss  \& Mebold (1977)  observed in  the direction  of Cir  X--1 seven
times.  The first observation was  done when Cir X--1 was supposed not
emitting in  the radio  band (Spectrum I)  while three  radio spectra,
corresponding to  three distinct observations, were  averaged when the
radio flux of Cir X--1 was around 1 Jy; Spectrum I was subtracted from
the averaged spectrum.  Supposing that no contaminating variable radio
sources  were present  during  the observations,  the absorption  line
corresponding to the farthest \ion{H}{1}  cloud gives a lower limit to
the distance to Cir X--1.   Goss \& Mebold (1977), assuming a distance
to the  center of  our Galaxy R$_0  = 10$  kpc and looking  toward the
direction  of  Cir  X--1  at  $l=322.117^\circ$  and  $b=0.038^\circ$,
obtained that the minimum distance between the Galactic center and the
tangential point $R=  R_0 \sin l$ is 6.1 kpc and  that the distance to
the tangential  point $d= R_0 \cos  l$ is 7.9 kpc.   At the tangential
point the radial velocity will be  maximum, in our case $v_r = -88 \pm
1$  km/s (see  Figure 2  in  Rohlfs et  al. 1986).   Since the  radial
velocity at  -75 km/s  extends up to  -90 km/s  and since -90  km/s is
compatible  with the  radial  velocity at  the  tangential point  they
assumed  $d= 7.9$ kpc  as lower  limit to  the distance  to Cir  X--1. 
Rescaling  these  results using  the  most  recent  estimation of  the
distance  to  the  Galactic center  $R_0  =  8.5  \pm 0.5$  (Feast  \&
Whitelock, 1997), the  lower limit to the distance  of Cir X--1 should
be $d= 6.7 \pm 0.4 $ kpc. However we note that: a) although the radial
velocity at -75  km/s extends down to -90  km/s the associated optical
depth  drastically  decreases  below  1  at  -76  km/s  becoming  less
prominent of  the absorption  line at 3  km/s (see  fig. 1 in  Goss \&
Mebold 1977); b) the absorption line  at 3 km/s is produced by a cloud
having a  distance larger than  the distance to the  tangential point,
implying  that the  lower limit  to the  Cir X--1  distance  should be
larger ($>  13$ kpc, see Tab.  \ref{tabdistances})  than that actually
accepted.   Note  that the  corresponding  equivalent hydrogen  column
associated  to  a  distance of  13  kpc  is  $\sim 3  \times  10^{22}$
cm$^{-2}$,  never observed  in the  X-ray  spectra of  Cir X--1.  This
suggests that possible contaminating  radio sources are present during
the observation. For these reason we analysed all the absorption lines
present in  the \ion{H}{1} spectrum.  To  do it we  used the following
relation:
$$
v_r=W(R) \sin l \cos b 
$$ 
$$
R^2=R_0^2+d^2-2R_0d\cos l
$$
(see Olling  \&  Merrifield, 1998  and  references therein).   The
function  $W(R)$  is  defined as  $W(R)=R_0(\Omega(R)-\Omega_0(R_0))$,
where $\Omega(R)$ is the angular  speed of a celestial body around the
Galactic   center,  $R$  the   radius  of   the  circular   orbit  and
$\Omega_0(R_0)$ is the angular speed of the local system of rest (LSR).
 Using  the method of the  tangential point, it  is possible to
write  the  function  $W(R)$  as  function  of $R$  (see  fig.   2  in
Merrifield, 1992  and references  therein); solving the  two equations
above we find the distances  reported in Tab.  \ref{tabdistances}.

In  order  to  discriminate  between  these  values  we  computed  the
corresponding visual extinction (Hakkila et al., 1997), the equivalent
hydrogen column (Predehl \& Schmitt  1995) and the proper motion $\mu$
(not corrected for the Solar  motion) for each distance.  These values
are also reported in Tab. \ref{tabdistances}.  Excluding the distances
of 0.5  and 0.9  kpc, because the  corresponding $N_H$ value  is lower
than that  observed in  the X--ray  spectra of Cir  X--1, we  can give
$d=4.1 \pm 0.3$ kpc as a lower  limit to the distance of Cir X--1.  On
the other  hand we can  exclude the distances  12.5 kpc, 12.9  kpc and
13.6 kpc because the corresponding  $N_H$ values are larger than those
observed in  the X--ray spectra of  Cir X--1. Using  this argument the
distance of  the source should be  in the range between  $4.1 \pm 0.3$
and  $9.3 \pm  0.6$  kpc.  Our  conclusion  is in  agreement with  the
results of  $d>4$ kpc obtained by  Whelan et al. (1977)  using data in
the optical band.  Finally we note that a) a recent HST observation of
the optical counterpart  of Cir X--1 indicates that  its proper motion
(not corrected  for the  Solar motion) is  $\mu =1.5 \pm  1.6$ mas/yrs
(Mignani et al.,  2002) similar to the proper  motion obtained for the
lower limit of  $d=4.1 \pm 0.3$ kpc (see  Tab. \ref{tabdistances}); b)
the value of $N_H$ that we obtain  from the fit in this work is around
$0.6  \times  10^{22}$   cm$^{-2}$  (see  Tabs.  \ref{tabnifeapo}  and
\ref{tabnifeperi}), compatible  with the  value of $N_H$  obtained for
$d=4.1 \pm 0.3$ kpc.  For this  reason. in the following we will adopt
a distance  to Cir X--1 of  4.1 kpc, the corresponding  value of $N_H$
will be $0.66 \times  10^{22}$ cm$^{-2}$ (see Tab. \ref{tabdistances})
that is the same fixed  in the fits reported in Tabs.  \ref{tabapofix}
and \ref{tabperifix}.

This value  of distance  to Cir X--1  slightly changes   the parameters
associated to the  radio jet observed from the  source (Fender et al.,
2004).  From  three radio observations Fender et  al.  (2004) resolved
the structure  of a jet  in Cir  X--1; the jet  is extended up  to 2.5
arcsec from the core.  The  authors, assuming a distance to the source
of 6.5  kpc, find a  superluminal motion of  the jet with  an apparent
velocity of $v_{app}  \simeq 15 c$; because of the  large value of the
apparent  velocity,  the angle  between  the  line  of sight  and  the
direction  of the jet,  $\theta$, is  $ <  5^\circ$ and  the intrinsic
velocity of the jet is $\beta >  0.998$. For a distance to Cir X--1 of
4.1  kpc, as  we suggest,  these  values slightly  change because  now
$v_{app}  \simeq 9.2  c$ implying  $\theta  < 12^\circ$  and $\beta  >
0.995$. The  extension of  the jet of  2.5 arcsec corresponds  to 0.12
light-years for 4.1 kpc, i.e.  $1.1 \times 10^{17}$ cm.  The detection
of the  superluminal motion in the  jet of Cir X--1  implies $\theta <
12^\circ$  and,   since  the  jet  should  have   a  direction  almost
perpendicular to the accretion disk,  it confirms that Cir X--1 is not
an edge-on  source.

At  the light  of these  results the  mechanism producing  the P-Cygni
profiles from  strongly ionized element,  observed by Brandt  \& Shulz
(2000) using  a  Chandra  observation,  should  be  rediscussed.   The
authors proposed a  radiatively driven wind along the  disk surface to
explain the presence  of these features in the  Chandra grating spectra
(see Brandt \&  Shulz, 2000; Shulz \& Brandt  2002 for more details
about their  proposed model)  but this mechanism  could be  valid only
assuming that Cir X--1 is an edge-on source.

A possibility that should be investigate is that the ionized matter is
ejected  in a  nearly perpendicular  direction to  the  accretion disk
surface.
 
\subsection{The absorption features in the spectra and possible Reflection} 

In this section  we discuss the absorption features  reported in Tabs. 
\ref{tabapofix} and \ref{tabperifix} and  obtained fixing the value of
$N_H$ to $0.66 \times 10^{22}$ cm$^{-2}$.

For the spectra from 1 to 7 
the absorption edges associated to \ion{O}{7}, \ion{O}{8}, \ion{Ne}{9}
are  probably  produced in  a  plasma  having  a ionization  parameter
log$_{10}\xi \sim 1.3$, for this  value of $\xi$ the fractional number
$f$  of  \ion{O}{7},  \ion{O}{8},  \ion{Ne}{9}  of  the  corresponding
elements  is  largest.   We  obtain  the  equivalent  hydrogen  column
corresponding  to the  optical  depth of  the  absorption edges  using
$\tau_X=   \sigma_X   N_{H}   A_X   f$,  where   $\sigma_X$   is   the
photoionization  cross   section  for  the  ion  X,   $N_{H}$  is  the
corresponding equivalent  hydrogen column,  $A_X$ is the  abundance of
the  element X  and $f$  the fractional  number of  ions in  the given
ionization  state  of  the  considered  element.   The  value  of  the
photoionization  cross sections  are $2.4  \times  10^{-19}$ cm$^{2}$,
$9.9 \times  10^{-20}$ cm$^{2}$ and  $1.5 \times 10^{-19}$ for  O {\sc
  vii}, \ion{O}{8}  and \ion{Ne}{9}, respectively (see  Verner et al.,
1996).  Using the values  of $\tau$ reported in Tab.  \ref{tabapofix},
the cosmic  abundance of oxygen  and neon, and assuming  an ionization
parameters  of  log$_{10}\xi  =   1.3$  (implying  $f  \sim  0.5$  for
\ion{O}{7}, \ion{O}{8}, \ion{Ne}{9}) we  find that N$_{H_{O\; VII}}$ $
\sim 3  \times 10^{22}$ cm$^{-2}$,  N$_{H_{O\; VIII}}$ $\sim  4 \times
10^{22}$  cm$^{-2}$  and N$_{H_{Ne\;  IX}}$  $\sim  2 \times  10^{22}$
cm$^{-2}$.  We also observe  an absorption edge probably associated to
\ion{Fe}{25}  at   9  keV.  The  photoionization   cross  sections  of
\ion{Fe}{25}  is $2.0 \times  10^{-20}$ cm$^{2}$  (see Verner  et al.,
1996).   Assuming a  ionization parameter  log$_{10}\xi$ of  3.3 (that
implies the largest value of  $f\sim 0.65$) we obtain that N$_{H_{Fe\;
    XXV}}$ is $\sim 8.3 \times 10^{22}$ cm$^{-2}$.  In spectra 6 and 7
an  emission line associated  to \ion{Fe}{25}  is present.   From $\xi
=L_x/n_e r^2$ (see Krolik, McKee  \& Tarter, 1981) and $L_{line}=4 \pi
D^2 I_{line}=n_e^2  V \alpha  A f$, we  can obtain the  distance, $r$,
from the central source and the corresponding electron density, $n_e$,
of the  region where the line  is produced.  In the  formulas $L_x$ is
the total unabsorbed x-ray luminosity of the source in the energy band
0.12--200 keV, $L_{line}$ the luminosity of the line, $D$ the distance
to the source, $I_{line}$ the  intensity of the line, $V$ the emitting
volume, $\alpha$ the recombination parameter, $A$ the cosmic abundance
of  iron.  We  assume a  spherical volume  $V$ of  radius $r$  and, as
written   above,  log$_{10}\xi=3.3$  implies   $f  \sim   0.65$.   The
recombination parameter  $\alpha$ is  obtained using the  relation and
the best-fit parameters for H-like and He-like ions reported by Verner
\&  Ferland (1996),  where  we  fixed the  plasma  temperature at  the
electron  temperature of  the  Comptonizing cloud;  we  find that  the
\ion{Fe}{25} line is produced at $r  \sim 2 \times 10^{10}$ cm, with a
corresponding  electron  density  of   $n_e  \sim  7  \times  10^{13}$
cm$^{-3}$.  We can also estimate an upper limit to the distance to the
emitting source,  $d$, of  the region where  the absorption  edges are
produced.  To do it we use $d<L_x/(N \xi)$ (Reynolds \& Fabian, 1995),
where  $N$  is  the  equivalent  hydrogen  column  associated  to  the
photoionized matter, $d$ the distance to the central object, and $\xi$
and  L$_x$  are  defined  as  above.  We  find  that  the  \ion{O}{7},
\ion{O}{8} and \ion{Ne}{9} absorption edges are produced at a distance
to  the  central object  of  $d  < 5.3  \times  10^{13}$  cm, and  the
\ion{Fe}{25} absorption  edge at  $d < 2.5  \times 10^{11}$  cm.  This
latter  distance is  compatible with  the distance  associated  to the
\ion{Fe}{25} absorption line.

In spectra  A and B we fixed  four absorption edges below  2 keV.  The
absorption edges  associated to  \ion{O}{7},  \ion{O}{8}  should be
produced in  a region having  a lower ionization parameter  (we assume
log$_{10}\xi=1$) with respect to the region where the absorption edges
of \ion{Mg}{11}    and   \ion{Mg}{12} are  produced   (we  assume
log$_{10}\xi=2.2$ for this latter region).  For the spectra A and B we
use a lower value of log$_{10}\xi=1$ associated to \ion{O}{7}, \ion{O}{8}
 than in the spectra from 1  to 7 because now we do not observe
the \ion{Ne}{9} absorption edge.

Using  the values of  $\tau$ reported  in tab.   \ref{tabperifix}, the
solar  abundance  of  oxygen   and  assuming  a  ionization  parameter
log$_{10}\xi$ of 1 (implying $f \sim  0.5$ for \ion{O}{7} and $f \sim
0.25$ for \ion{O}{8} we find that  N$_{H_{O\; VII}}$ and N$_{H_{O\;
    VIII}}$ are $ \sim 4  \times 10^{22}$ cm$^{-2}$.  In the same way,
knowing that  the values of  the photoionization cross sections  of 
\ion{Mg}{11} and \ion{Mg}{12}  are $1.0 \times 10^{-19}$ cm$^{2}$ and $4.4
\times 10^{-20}$  cm$^{2}$ respectively (see Verner et  al., 1996) and
assuming a ionization parameter log$_{10}\xi$ of 2.2 (that implies the
largest values of $f\sim 0.4$ for  \ion{Mg}{11} and $f\sim 0.45$ for 
\ion{Mg}{12}),  we obtain that  N$_{H_{Mg\; XI}}$ and  N$_{H_{Mg\; XII}}$
are $\sim  7 \times 10^{22}$  cm$^{-2}$.  An absorption  edge probably
associated  to  \ion{Fe}{23}  is  also present  at  8.5  keV.  The
photoionization  cross  sections of \ion{Fe}{23}   is $2.2  \times
10^{-20}$ cm$^{2}$ (Krolik \&  Kallman, 1987).  Assuming an ionization
parameter  log$_{10}\xi$ of  2.9 (that  implies the  largest  value of
$f\sim 0.25$)  we obtain  that N$_{H_{Fe\; XXIV}}$  is $\sim  4 \times
10^{24}$ cm$^{-2}$, as already proposed by Iaria et al.  (2001a).  The
upper limit to  the distance of the region  where this absorption edge
is produced is $d < 3.3 \times 10^{10}$ cm.

Two  absorption lines, associated  to  \ion{O}{8}  and  \ion{Mg}{12}
respectively, are  present.  The ionization  parameter, log$_{10}\xi$,
associated to  \ion{O}{8}  is 1 implying  $f \sim 0.25$.   We obtain
that  the \ion{O}{8} absorption line  is produced  at $r  \sim 1
\times 10^{13}$ cm, with a corresponding electron density of $n_e \sim
9   \times   10^{10}$  cm$^{-3}$.    The   ionization  parameter   ,
log$_{10}\xi$,  associated to   \ion{Mg}{12} is  2.2 implying  $f \sim
0.45$; we find that the  \ion{Mg}{12} absorption line is produced at $r
\sim 1 \times 10^{11}$ cm,  with a corresponding electron density of
$n_e \sim  6.2 \times 10^{13}$ cm$^{-3}$.   On the other  hand we find
that the \ion{O}{7} and \ion{O}{8} absorption edges are produced at
$d  < 2.5  \times  10^{14}$  cm, the   \ion{Mg}{11}  and   \ion{Mg}{12}
absorption edges  are produced  at $d <  1 \times 10^{13}$  cm.  These
upper limits  are compatible  with the distances  derived for  the 
\ion{O}{8} and  \ion{Mg}{12} absorption lines.

Finally an emission line at  6.7 keV is observed. Under the hypothesis
that this  is produced by \ion{Fe}{25} we find that  the distance and
the electron density  of the region producing this line  are $r \sim 6
\times  10^{11}$  cm  and  $n_e  \sim 1.6  \times  10^{11}$  cm$^{-3}$
respectively.  This result is unrealistic because implies the presence
of  plasma with a  ionization parameter  of log$_{10}\xi  = 3.3$  at a
larger distance than the plasma  having log$_{10}\xi = 2.2$, where the
\ion{Mg}{12} absorption line  is produced. A possible  explanation is
that the  emission line  at $\sim  6.7$ keV is  associated to \ion{Fe}{23};
 the corresponding centroid of the line should be $\sim 6.62$
keV and this does not change the results of the best fit for spectra A
and B.  In this  case unfortunately we  do not know  the recombination
parameter  of  the line  and,  consequently  we  cannot calculate  the
distance to  the region where this  line is produced. 

Summarizing  these results  indicate  that ionized  matter is  present
around the  system. During the  observation at phases  0.62--0.84 this
ionized matter could have an electron density quite constant along the
distance of  $\sim 7\times 10^{13}$ cm$^{-3}$;  during the observation
at  phases 0.11--0.16  the electron  density decreases  going  from the
compact object to larger distances, varying between 9 $\times 10^{10}$
cm$^{-3}$ at 1 $\times 10^{13}$  cm and 6 $\times 10^{13}$ cm$^{-3}$
at 1 $\times 10^{11}$ cm. 
 The different distribution of
ionized matter along the distance from the neutron star, during the two
observations, may influence the different observed continuum emission
(see  section \ref{cont}).  
In Tab. \ref{tabioni} we summarize the obtained results.

Finally we  observe an absorption  edge associated to neutral  iron in
all  the nine  spectra.  This  feature  was also  observed during  the
Chandra observations (Brandt \& Shulz, 2000; Shulz \& Brandt, 2002).
The optical depth is $\sim 0.05$ for spectra from 1 to 7 and less than
$ 0.09$  in spectra A  and B.  Knowing  that the cross section  of the
iron  K-edge  is $3.7  \times  10^{-20}$  cm$^{2}$  we find  that  the
equivalent hydrogen column associated  to this feature is N$_{H_{Fe \;
    I}}  \sim 3  \times 10^{22}$  cm$^{-2}$  and N$_{H_{Fe  \;I}} <  5
\times 10^{22}$ cm$^{-2}$ for the observation at phases 0.62--0.84 and
0.11--0.16,  respectively. The  estimated  equivalent hydrogen  column
associated to  the neutral iron is  about 5 times higher  than seen in
the  low-energy  absorption indicating  an  overabundance  of iron  or
special geometrical conditions. A  plausible geometry was suggested by
Singh  \& Apparao  (1994) which  found similar  results for  the Atoll
source 4U  0614+09 analyzing  EXOSAT data. The  \ion{Fe}{1} absorption
edge  is  imprinted on  the  spectrum due  to  reflection  by cold  or
partially ionized matter which is not in the same line of sight of the
direct  emission, therefore,  lead to  an estimate  of N$_H$  which is
different from the value obtained from low-energy absorption.

\subsection{The parameters of the continuum emission components}
\label{cont}
The  model used  to fit  the data  taken at  the phases  0.62--0.84 is
typical  of the  Z-sources  (see  Iaria et  al.,  2003 and  references
therein).  It is composed by a blackbody plus a Comptonized component,
and a power-law at energies above 10 keV.  The corresponding radius of
the blackbody component, assuming  a spherical emission and a distance
of 4.1 kpc is $\sim 20$ km.  This radius is too large to be associated
with  the  emission  from  the  neutron star  surface;  therefore  the
blackbody  component is  probably emitted  by  the inner  part of  the
accretion disk. Following in 't  Zand et al.  (1999) we calculated the
radius, $R_W$, of the seed-photon emitting region using the parameters
reported in Table \ref{tabapofix} and  a distance to the source of 4.1
kpc; we obtain $R_W \sim 6$  km; this is smaller than the neutron star
radius;  a possible  implication is  that  the electron  cloud is  not
spherical.   The  continuum  obtained  at the  phases  0.11--0.16  was
already discussed  by Iaria et  al.  (2001a), it  is fitted by  a very
soft Comptonized  component plus  a power-law component,  no blackbody
emission is observed.   In this case $R_W$ is $\sim  100$ km.  In this
case the shielding of the inner regions could be produced by a thicker
optically region  of the  corona at inner  radius.  Assuming  that the
Comptonizing corona  produces also the absorption  feature observed in
the spectra we deduce that during the observation at phases 0.62--0.84
the ionized matter has a low  density allowing us to observe the inner
region of the system up to 20 km from the compact object, while during
the observation at phases 0.11-0.16  the density of the ionized matter
is highest  near the  compact object forming  a curtain that  does not
allow us to observe the inner region.

We also  note that,  although in the  two observations  considered in
this  paper the  continuum emission  below  10 keV  is different,  the
spectra above 10 keV are  fitted by a power-law component described by
the same  parameters, suggesting that  its presence is  independent of
the phase and  luminosity of the X-ray source.   As described in other
works  (see Iaria et  al., 2003  and references  therein) it  could be
connected to the  presence of a motion of  relativistic electrons in a
jet, as also suggested by Fender et al.  (2004, see below).
 
The  intrinsic total  luminosity  in the  energy  band 0.12--200  keV,
L$_x$, is $\sim 4 \times  10^{37}$ erg s$^{-1}$ during the observation
at phases 0.62--0.84 and $\sim 1 \times 10^{38}$ erg s$^{-1}$ during
the observation at  phases 0.11--0.16 for a distance  to the source of
4.1 kpc.

\section{Conclusions}

In this work we present the analysis of the broad band (0.12--200 keV)
spectrum of  Cir X--1 at phases  0.62--0.84 and the  reanalysis of the
broad band (0.12--200 keV) spectrum  of Cir X--1 at phases 0.11--0.16,
using   two  BeppoSAX  observations   taken  in   2001  and   1998  in
respectively.  In the  spectra, a soft excess below  1 keV is observed
using the model previously proposed.   We fitted the soft excess using
the equivalent  hydrogen column of 0.66 $\times  10^{22}$ cm$^{-2}$ and
adding at low energies absorption edges  of \ion{O}{7}, \ion{O}{8},
\ion{Ne}{9}, \ion{Mg}{11} and \ion{Mg}{12}. Moreover in the observation
at phases 0.11-0.16  two absorption lines of \ion{O}{8} and \ion{Mg}{12}
 were added.

The  equivalent hydrogen column  of the  interstellar matter  of $0.66
\times 10^{22}$  cm$^{-2}$ gives a  distance to Cir  X--1 of 4.1  kpc. 
This result is  discussed comparing it to the  values obtained by Goss
\& Mebold (1977)  and reanalyzing the HI spectrum  of the source. 

We discuss the different continuum observed in the two observations as due
to  the different  density of  the  ionized matter  around the  binary
system in the two observations.
 
Finally the reader should  note that our results could be model dependent; 
for this reason further spectroscopic analysis of the Cir X--1 spectrum 
below 1 keV are need to confirm the scenario proposed in this work.

\acknowledgments This work  was partially  supported by  the Italian
Space  Agency   (ASI)  and  the  Ministero   della  Istruzione,  della
Universit\'a e della Ricerca (MIUR).


\clearpage

\begin{deluxetable}{ccccccc}
\tablewidth{0pt} 
\tablecaption {  \label{tab1} BeppoSAX  observation at
    phases 0.62-0.84.   In columns  1 and 2  we show the  intervals of
    Soft Color (SC) and Hard Color (HC) which identify the seven zones
    selected  in  the CD.   In  the  other  columns the  corresponding
    exposure   time,  in  ks,   for  each   instrument  is   shown.  }
\tablewidth{0pt} 
\tablehead{ 
\colhead{ } 
&\colhead{SC} 
&\colhead{HC}
&\colhead{LECS} 
&\colhead{MECS} 
&\colhead{HP}
&\colhead{PDS} \\

\colhead{ }
&\colhead{Interval}
&\colhead{Interval}
&\colhead{ks}
&\colhead{ks}
&\colhead{ks}
&\colhead{ks}
}
\startdata
  1 & 2.24-2.30 & 0.145-0.15 & 4.7 & 8.7 & 7.0 & 3.4 \\
  2 & 2.20-2.26 & 0.139-0.145 & 7.7 & 32.0 & 28.3 & 14.1 \\
  3 & 2.18-2.24 & 0.135-0.139 & 9.7 & 32.1 & 47.6 & 14.9 \\
  4 & 2.16-2.22 & 0.132-0.135 & 2.0 & 8.9 & 8.8 & 4.3 \\
  5 & 2.08-2.2 & 0.124-0.132 & 1.9 & 11.9 & 11.4 & 4.7 \\
  6 & 2.24-2.26 & 0.124-0.176 & 1.9 & 5.1 & 4.0 & 1.4 \\
  7 & 2.6-2.9 & 0.124-0.176 & 4.8 & 10.5 & 8.6 & 9.9 \\
\tableline
\enddata 
\end{deluxetable}

\begin{deluxetable}{ccccccc}
\tablecaption { \label{tab2} The same as in Tab. \ref{tab1} for the BeppoSAX 
observation at  phases  0.11-0.16. }
\tablewidth{0pt}
\tablehead{
\colhead{$~$ }
&\colhead{SC}
&\colhead{HC}
&\colhead{LECS}
&\colhead{MECS}
&\colhead{HP}
&\colhead{PDS} \\ 
\colhead{$~$ }
&\colhead{Interval}
&\colhead{Interval}
&\colhead{ks}
&\colhead{ks}
&\colhead{ks}
&\colhead{ks}
}
\startdata
 A & 0.52-0.565 & 0.012-0.015 & 2.5 & 5.0  &  4.8 & 5.0 \\
 B & 0.565-0.64 & 0.014-0.019 & 1.7 & 9.8  & 11.4 & 11.1 \\
 C & 0.68-0.71 & 0.02-0.024 & 0 & 0.4  & 1.0  & 1.0 \\
\tableline
\enddata 
\end{deluxetable}

\begin{table}
\scriptsize
\begin{center}
\caption{ \footnotesize \label{Tabbaseapo}
  Results of the spectral fits in the 
  0.12--200 keV energy band for the seven spectra corresponding to the
  observation  taken  at  the  phases 0.62--0.84.   The  photoelectric
  absorption is indicated as $N_{\rm  H}$.  The model is composed by a
  blackbody   plus  a   Comptonized  spectrum   modeled  by   Comptt.  
  Uncertainties are  at 90\% confidence level for  a single parameter,
  upper  limits  are at  95\%  confidence  level.   $kT_{\rm BB}$  and
  N$_{\rm  BB}$  are,  respectively,  the  blackbody  temperature  and
  normalization in  units of $L_{39}/D_{10}^2$, where  $L_{39}$ is the
  luminosity in units of $10^{39}$ ergs/s and $D_{10}$ is the distance
  in  units  of  10  kpc.   kT$_0$, kT$_e$  and  $\tau$  indicate  the
  seed-photon  temperature, the electron  temperature and  the optical
  depth  of the Comptonizing  cloud around  the neutron  star. N$_{\rm
    comptt}$  is  the  normalization  of  the Comptt  model  in  XSPEC
  v.11.2.0 units. }
\begin{tabular}{l|c|c|c|c|c|c|c} 
\hline
\hline
Parameters &  1 &2 &3 &4 & 5 & 6& 7         \\
\hline
$N_{\rm H}$ $\rm (\times 10^{22}\;cm^{-2})$ 
& $1.644^{+0.051}_{-0.049}$ 
& $1.725^{+0.041}_{-0.040}$ 
& $1.703^{+0.037}_{-0.036}$ 
& $1.750^{+0.074}_{-0.070}$ 
& $1.791^{+0.077}_{-0.072}$ 
& $1.730^{+0.077}_{-0.074}$ 
& $1.689^{+0.049}_{-0.048}$ \\
&&&&&&&\\
kT$_{BB}$ (keV)
& $0.549 \pm 0.025$ 
& $0.531 \pm 0.017$ 
& $0.537 \pm 0.015$ 
& $0.523 \pm 0.024$  
& $0.505 \pm 0.020$ 
& $0.520 \pm 0.022$ 
& $0.550 \pm 0.017$ \\

N$_{BB}$   $(\times 10^{-2})$ 
& $4.67 \pm 0.28$ 
& $4.77 \pm 0.18$ 
& $4.85 \pm 0.17$ 
& $4.87 \pm 0.29$ 
& $4.60 \pm 0.26$ 
& $4.51 \pm 0.25$ 
& $4.68 \pm 0.17$ \\
&&&&&&&\\

 kT$_{0}$ (keV) 
& $1.035 \pm 0.040$ 
& $1.034 \pm 0.025$
& $1.060 \pm 0.023$
& $1.019 \pm 0.032$ 
& $1.040 \pm 0.030$ 
& $1.187 \pm 0.037$
& $1.251 \pm 0.028$\\
  
 kT$_{e}$ (keV) 
& $2.820^{+0.038}_{-0.035}$ 
& $2.796^{+0.026}_{-0.024}$ 
& $2.725^{+0.026}_{-0.025}$ 
& $2.751^{+0.038}_{-0.036}$ 
& $2.704^{+0.061}_{-0.056}$ 
& $2.678^{+0.091}_{-0.080}$ 
& $2.715^{+0.058}_{-0.052}$ \\

 $\tau$ 
& $11.93 \pm 0.30$ 
& $11.56 \pm 0.21$ 
& $11.28 \pm 0.22$ 
& $11.36 \pm 0.29$ 
& $10.70 \pm 0.39$ 
& $10.30 \pm 0.62$ 
& $10.43 \pm 0.44$\\ 

N$_{comptt}$
& $1.071 \pm 0.041$ 
& $1.074^{+0.028}_{-0.027}$
& $1.088^{+0.028}_{-0.026}$ 
& $1.127^{+0.040}_{-0.041}$ 
& $0.991 \pm 0.041$ 
& $1.109^{+0.054}_{-0.057}$
& $1.256^{+0.039}_{-0.041}$ \\
&&&&&&&\\

$\chi^2$ (d.o.f.)
& 285 (202) 
& 364 (200) 
& 330 (200) 
& 254 (201) 
& 244 (199) 
& 249 (199) 
& 256 (200) \\
\tableline
\end{tabular}
\end{center}
\end{table}

\begin{table}
\scriptsize
\begin{center}
\caption{\footnotesize Parameters of the best fit model in the 
  0.12--200 keV energy band for the seven spectra corresponding to the
  observation taken at the  phases 0.62--0.84.  The continuum emission
  is fitted by a blackbody plus a Comptonized component (Comptt) and a
  power-law component.  Uncertainties are at 90\% confidence level for
  a single parameter, upper limits  are at 95\% confidence level.  The
  power-law normalization, N$_{po}$, is in units of photons keV$^{-1}$
  cm$^{-2}$ s$^{-1}$ at 1  keV.  E$_{edge_1}$ and E$_{edge_2}$ are the
  energies   of  the   two  absorption   edges,   $\tau_{edge_1}$  and
  $\tau_{edge_2}$  their  corresponding  optical  depths (see  text).  
  E$_{\rm Fe \;{I}}$ and $\tau_{\rm  Fe \;{I}}$ are the energy and the
  corresponding optical depth of the  edge associated to neutral iron. 
  E$_{\rm Fe \;{XXV}}$ and $\tau_{\rm  Fe \;{XXV}}$ are the energy and
  the corresponding  optical depth of  the edge associated to  Fe {\sc
    xxv}.  E$_{\rm  Fe \;{\rm  XXV}}$, $\sigma_{\rm Fe  \;{\rm XXV}}$,
  I$_{\rm  Fe  \;{\rm  XXV}}$  and  EQW$_{\rm Fe  \;{\rm  XXV}}$  are,
  respectively, the centroid, width, intensity and equivalent width of
  the emission line at $\sim 6.7$ keV.  I$_{\rm Fe \;{\rm XXV}}$ is in
  units  of  photons cm$^{-2}$  s$^{-1}$.   The  other parameters  are
  defined as in Tab.  \ref{Tabbaseapo}.}
\begin{tabular}{l|c|c|c|c|c|c|c} 
\hline
\hline
Parameters &  1 &2 &3 &4 & 5 & 6& 7         \\
\hline
$N_{\rm H}$ $\rm (\times 10^{22}\;cm^{-2})$ 

& $0.57 \pm 0.12$  
& $0.605^{+0.113}_{-0.085}$    
& $0.67 \pm 0.13 $   
& 0.60 (fixed)
& 0.60 (fixed)
& 0.60 (fixed)
& $0.45^{+0.16}_{-0.11}$   \\
& & & & & & &\\

E$_{edge_1}$  (keV)
& $0.697^{+0.044}_{-0.031}$  
& $0.697^{+0.045}_{-0.032}$  
& $0.724^{+0.051}_{-0.041}$  
& $0.680^{+0.046}_{-0.064}$   
& $0.694^{+0.045}_{-0.079}$ 
& $0.650^{+0.065}_{-0.430}$
& $0.659^{+0.044}_{-0.039}$ \\

$\tau_{edge_1}$ 
& $5.6 \pm 1.6$  
& $5.2 \pm 1.8$    
& $5.13^{+1.71}_{-0.71}$  
& $8.0^{+2.9}_{-1.7}$ 
& $7.4^{+2.4}_{-1.5}$ 
& $5.4^{+2.8}_{-2.1}$
& $6.3^{+2.9}_{-2.0}$ \\
& & & & & & &\\

E$_{edge_2}$ (keV) 
& $1.006 \pm 0.079$  
& $0.955^{+0.096}_{-0.074}$  
& $1.025^{+0.177}_{-0.084}$  
& $1.209 \pm 0.078$ 
& $1.247\pm 0.079$  
& $0.962^{+0.079}_{-0.088}$  
& $0.978 \pm 0.053$ \\

$\tau_{edge_2}$ 
&  $0.90^{+0.78}_{-0.47}$  
& $1.23^{+1.08}_{-0.78}$  
& $0.66^{+0.75}_{-0.49}$     
& $0.38^{+0.20}_{-0.16}$      
& $0.28 \pm 0.16$   
&   $1.32^{+1.10}_{-0.85}$   
& $1.30^{+0.67}_{-0.52}$       \\
& & & & & & &\\

E$_{\rm Fe \;{\rm I}}$ (keV)
&  $7.08 \pm 0.11$ 
&  $7.217^{+0.096}_{-0.074}$   
& $7.127^{+0.093}_{-0.087}$ 
& $7.08 \pm 0.13$ 
&  $7.103^{+0.096}_{-0.092}$  
&  7.117 (fixed)
&  7.117 (fixed)  \\

$\tau_{\rm Fe \;{\rm I}}$ 
& $0.063 \pm 0.019$  
& $0.056 \pm 0.013$   
& $0.061 \pm 0.013$ 
& $0.055 \pm 0.017$    
& $0.076 \pm 0.017$   
& $0.031 \pm 0.029$
& $0.035 \pm 0.020$  \\
& & & & & & &\\

E$_{\rm Fe \;{\rm XXV}}$ (keV) 
& $9.12 \pm 0.21$  
& $8.96 \pm 0.26$ 
& $9.08^{+0.37}_{-0.28}$ 
& $8.80 \pm 0.18$   
& $8.96 \pm 0.23$  
&  $9.04 \pm 0.48$
& $8.91^{+0.28}_{-0.24}$   \\

$\tau_{\rm Fe \;{\rm XXV}}$ 
& $0.053 \pm 0.019$  
& $0.051 \pm 0.013$ 
& $0.038 \pm 0.013$   
& $0.064 \pm 0.017$   
& $0.071 \pm 0.019$  
& $0.036 \pm 0.025$ 
& $0.038 \pm 0.017$  \\
& & & & & & &\\

E$_{ \rm Fe \;{\rm XXV}}$ (keV)
&  --
& --
&  --
&  --
&  --
& $6.67^{+0.20}_{-0.11}$
& $6.740^{+0.093}_{-0.051}$\\

$\sigma_{ \rm Fe \;{\rm XXV}}$ (keV)
&  --
& --
&  --
&  --
&  --
& $<0.58$
& $<0.23$\\

I$_{\rm  Fe \;{\rm XXV}}$   ($\times 10^{-3}$) 
&  -- 
&  --
&  --
&  --
&  -- 
&  $4.5^{+2.5}_{-2.1}$ 
&  $3.2^{+1.7}_{-1.2}$  \\
& & & & & & &\\

kT$_{BB}$ (keV)
& $0.642^{+0.048}_{-0.053}$ 
& $0.617 \pm 0.037$  
& $0.611 \pm 0.032$  
& $0.575 \pm 0.052$ 
& $0.567 \pm 0.044$  
& $0.561 \pm 0.052$ 
& $0.613 \pm 0.038$      \\

N$_{BB}$   $(\times 10^{-2})$ 
& $4.58 \pm 0.81 $ 
& $4.26^{+0.54}_{-0.50} $ 
& $4.47 \pm 0.40 $   
& $4.23^{+0.64}_{-0.55} $  
& $4.06^{+0.40}_{-0.36} $   
& $3.47^{+0.34}_{-0.39} $    
& $3.82 \pm 0.36$  \\
& & & & & & &\\

 kT$_{0}$ (keV) 
& $1.14 \pm 0.12$ 
& $1.099^{+0.073}_{-0.065}$ 
& $1.138 ^{+0.060}_{-0.055}$ 
& $1.048^{+0.088}_{-0.074}$ 
& $1.090^{+0.077}_{-0.068}$ 
& $1.183^{+0.097}_{-0.084}$ 
& $1.295 \pm 0.072$  \\
  
 kT$_{e}$ (keV) 
& $2.709^{+0.075}_{-0.066}$ 
& $2.656^{+0.044}_{-0.041}$  
& $2.624^{+0.047}_{-0.043}$ 
& $2.608^{+0.063}_{-0.057}$ 
& $2.578^{+0.073}_{-0.067}$ 
& $2.58 \pm 0.11$ 
& $2.664 \pm 0.090$   \\

 $\tau$ 
& $12.83 \pm 0.76$ 
& $12.69 \pm 0.45$ 
& $12.02 \pm 0.45$ 
& $12.58 \pm 0.60$ 
& $11.79 \pm 0.69$ 
& $11.2 \pm 1.1$ 
& $10.79 \pm 0.86$   \\

N$_{comptt}$
& $1.003^{+0.082}_{-0.077}$  
& $1.028 \pm 0.055$  
& $1.039 \pm 0.048$  
& $1.120 \pm 0.079$   
& $0.974 \pm 0.066$   
& $1.121^{+0.083}_{-0.086}$ 
& $1.227 ^{+0.069}_{-0.076}$  \\
& & & & & & &\\
 
Photon Index 
& 2.60 (fixed) 
& 2.60 (fixed) 
& 2.60 (fixed) 
& 2.60 (fixed)  
& 2.60 (fixed)
& 2.60 (fixed)
& 2.60 (fixed)  \\

N$_{po}$     
& $0.21 \pm 0.15$  
& $0.208 \pm 0.074$ 
& $0.180 \pm 0.065$ 
& $0.18  \pm 0.12$ 
& $< 0.05$
& $< 0.13$
& $< 0.16$  \\
& & & & & & &\\

$\chi^2$ (d.o.f.)
& 203 (193) 
& 246 (191) 
& 222 (191) 
& 189 (193) 
& 182 (191) 
& 198 (189) 
& 161 (189) \\
\hline     
\hline 
\end{tabular}
\label{tabnifeapo}
\end{center}
\end{table}

\begin{table}[t]
\begin{center}
\scriptsize
\caption{ \footnotesize  Parameters of the best fit model in the 
  0.12--200 keV  energy band for  the spectra A  and B.  The  model is
  composed  by  a  Comptonized  component (Comptt)  plus  a  power-law
  component.  Uncertainties are at  90\% confidence level for a single
  parameter,  upper   limits  are  at  95\%   confidence  level.   The
  parameters are defined as in Tab.  \ref{tabnifeapo}.  }
\begin{tabular}{l|c|c} 
\hline
\hline
Parameters 
& A
& B        \\
\hline
$N_{\rm H}$ $\rm (\times 10^{22}\;cm^{-2})$

& $0.597^{+0.091}_{-0.088}$ 
& $0.621^{+0.140}_{-0.078}$ \\
& & \\

E$_{edge_1}$ (keV) 
& $0.624^{+0.042}_{-0.103}$
& 0.624 (fixed) \\

$\tau_{edge_1}$  
& $4.3^{+3.3}_{-1.3}$ 
& $3.6 \pm 1.7$ \\
& & \\

E$_{edge_2}$ (keV) 
&  $0.824^{+0.039}_{-0.034}$
&  $0.809 \pm 0.034$   \\

$\tau_{edge_2}$ 
&  $2.61^{+0.91}_{-0.97}$
&  $2.60 \pm 0.81$    \\
& & \\

E$_{edge_3}$ (keV)
&  $1.352 \pm 0.029$
&  $1.303 \pm 0.045$    \\

$\tau_{edge_3}$ 
&  $0.145^{+0.048}_{-0.052}$  
&  $0.118^{+0.028}_{-0.053}$    \\
& & \\
E$_{edge_4}$ (keV) 
&  $1.883 \pm 0.025$
&  $1.870 \pm 0.028$    \\

$\tau_{edge_4}$ 
&  $0.128 \pm 0.023$
&  $0.122^{+0.024}_{-0.027}$    \\
& & \\

E$_{\rm Fe \;{\rm XXIII}}$ (keV)
& $8.452 \pm 0.058$
& $8.467 \pm 0.046$  \\

$\tau_{\rm Fe \;{\rm XXIII}}$ 
& $0.809^{+0.109}_{-0.088}$ 
& $0.709^{+0.125}_{-0.081}$   \\
& & \\

E$_{\rm Fe \;{\rm I}}$ (keV)
&  7.117 (fixed)
&  7.117 (fixed)  \\

$\tau_{\rm Fe \;{\rm I}}$
& $< 0.070$ 
& $<0.090$  \\
& & \\

E$_{ \rm Fe \;{\rm XXV}}$ (keV)
& 6.700 (fixed)
& 6.700 (fixed)\\

$\sigma_{\rm Fe\;{\rm XXV}}$ (keV)
&  $<0.25$
&   $<0.61$\\

I$_{\rm Fe\;{\rm XXV}}$  $(\times 10^{-3})$  
& $1.18^{+0.55}_{-0.72}$ 
& $1.86^{+1.42}_{-0.84}$\\
& & \\

 kT$_{0}$ (keV) 
& $0.416  ^{+0.018}_{-0.015}$
& $0.439 \pm 0.019$ \\
  
 kT$_{e}$ (keV) 
& $0.933^{+0.065}_{-0.028}$ 
& $0.997^{+0.091}_{-0.042}$ \\

 $\tau$ 
& $14.09^{+0.71}_{-1.37}$ 
& $13.7^{+1.1}_{-1.6}$  \\

N$_{comptt}$
& $32.5^{+1.9}_{-4.3}$ 
& $27.3 ^{+3.6}_{-4.7}$  \\
& & \\
 
Photon Index 
& 2.60 (fixed)
& 2.60 (fixed) \\

N$_{po}$     
& $0.327 \pm 0.063$  
& $0.228^{+0.054}_{-0.068}$   \\
 & &\\

$\chi^2$ (d.o.f.)
& 184 (169) 
& 214 (169) \\

\hline     
\hline 
\end{tabular}
\label{tabnifeperi}
\end{center}
\end{table}

\begin{table}
\scriptsize
\begin{center}
\caption{\footnotesize Parameters of the best fit model in the 
  0.12--200 keV energy band for the seven spectra corresponding to the
  observation taken  at  phases 0.62--0.84.  The continuum emission
  is the same as reported in table \ref{tabnifeapo}. The absorption edges
  at  low  energies are  interpreted  as  produced  by ions  of  light
  elements  and  their  corresponding  optical depths  are  reported.  
  Uncertainties are  at 90\% confidence level for  a single parameter,
  upper   limits   are  at   95\%   confidence  level.    Flux$_{BB}$,
  Flux$_{Comptt}$  and Flux$_{po}$  are  the intrinsic  fluxes in  the
  energy  band 0.12--200  keV,  in units  of  $10^{-8}$ erg  cm$^{-2}$
  s$^{-1}$, of the blackbody, of  the Comptonized component and of the
  power-law component, respectively.   L$_{tot}$ is the intrinsic total
  luminosity,  in the  energy  band  0.12--200 keV,  in  units of  erg
  s$^{-1}$ and assuming a distance to the source of 4.1 kpc.  }
\begin{tabular}{l|c|c|c|c|c|c|c} 
\hline
\hline
Parameters &  1 &2 &3 &4 & 5 & 6& 7         \\
\hline
$N_{\rm H}$ $\rm (\times 10^{22}\;cm^{-2})$ 

& 0.66 (fixed)
& 0.66 (fixed)
& 0.66 (fixed)
& 0.66 (fixed)
& 0.66 (fixed)
& 0.66 (fixed)
& 0.66 (fixed) \\
& & & & & & &\\

E$_{\rm O \; {\rm VII}}$ (keV) 
& 0.739 (fixed)
& 0.739 (fixed)
& 0.739 (fixed)
& 0.739 (fixed)
& 0.739 (fixed)
& 0.739 (fixed)
& 0.739 (fixed) \\

$\tau_{\rm O \; {\rm VII}}$ 
& $3.20^{+2.13}_{-0.93}$  
& $3.79^{+2.50}_{-0.93}$  
& $3.87^{+2.18}_{-0.92}$  
& $3.0^{+2.9}_{-1.1}$ 
& $5.54^{+0.46}_{-2.38}$ 
& $3.3^{+3.1}_{-1.3}$ 
& $2.39^{+1.23}_{-0.70}$ \\
& & & & & & &\\

E$_{\rm O \; {\rm VIII}}$ (keV) 
& 0.871 (fixed)
& 0.871 (fixed)
& 0.871 (fixed)
& 0.871 (fixed)
& 0.871 (fixed)
& 0.871 (fixed)
& 0.871 (fixed) \\

$\tau_{\rm O \; {\rm VIII}}$ 
& $1.63^{+0.57}_{-1.30}$   
& $<2.09$  
& $<1.90$  
& $<2.41$ 
& $<1.52$ 
& $<2.51$  
& $2.05^{+0.48}_{-0.78}$ \\
& & & & & & &\\

E$_{\rm Ne \; {\rm IX}}$ (keV)
& 1.196 (fixed)
& 1.196 (fixed)
& 1.196 (fixed)
& 1.196 (fixed)
& 1.196 (fixed)
& 1.196 (fixed)
& 1.196 (fixed) \\
 
$\tau_{\rm Ne \; {\rm IX}}$ 
&  $0.16 \pm 0.10$   
& $0.126 \pm 0.085$  
& $0.168 \pm 0.074$   
& $0.35 \pm 0.15$   
& $0.31 \pm 0.15$   
&  $<0.17$ 
& $<0.21$     \\
& & & & & & &\\

E$_{\rm Fe \;{\rm I}}$  (keV)
& 7.117 (fixed) 
& 7.117 (fixed) 
& 7.117 (fixed) 
& 7.117 (fixed) 
& 7.117 (fixed) 
& 7.117 (fixed) 
& 7.117 (fixed) \\

$\tau_{\rm Fe \;{\rm I}}$ 
& $0.059 \pm 0.018$  
& $0.052 \pm 0.013$   
& $0.060 \pm 0.013$ 
& $0.054 \pm 0.017$    
& $0.077 \pm 0.017$   
& $0.038 \pm 0.027$  
& $0.029 \pm 0.019$  \\
& & & & & & &\\

E$_{\rm Fe \;{\rm XXV}}$ (keV)
& $9.14 \pm 0.21$  
& $8.87^{+0.25}_{-0.20}$  
& $9.07^{+0.36}_{-0.29}$   
& $8.81 \pm 0.17$   
& $8.96 \pm 0.22$  
& $9.02 \pm 0.44$ 
& $8.91^{+0.29}_{-0.25}$    \\

$\tau_{\rm Fe \;{\rm XXV}}$ 
& $0.050 \pm 0.018$  
& $0.050 \pm 0.012$ 
& $0.036 \pm 0.013$   
& $0.061 \pm 0.017$   
& $0.071 \pm 0.019$  
& $0.040 \pm 0.025$ 
& $0.036 \pm 0.018$  \\
& & & & & & &\\

E$_{ \rm Fe \;{\rm XXV}}$ (keV)
&  --
& --
&  --
&  --
&  --
& 6.700 (fixed)
& 6.700 (fixed)\\

$\sigma_{ \rm Fe \;{\rm XXV}}$ (keV)
&  --
& --
&  --
&  --
&  --
& $<0.44$
& $<0.23$\\

I$_{\rm  Fe \;{\rm XXV}}$   ($\times 10^{-3}$) 
&  -- 
&  --
&  --
&  --
&  -- 
&  $4.2^{+2.3}_{-1.9}$ 
&  $3.3^{+1.7}_{-1.3}$  \\

EQW$_{ \rm Fe \;{\rm XXV}}$ (eV)
& --
& --
& --
& --
&--
& $30^{+17}_{-13}$ 
&$19^{+10}_{-7}$  \\
& & & & & & &\\

kT$_{BB}$ (keV)
& $0.623^{+0.049}_{-0.052}$ 
& $0.607 \pm 0.035$ 
& $0.604 \pm 0.029$  
& $0.584 \pm 0.048$ 
& $0.575 \pm 0.040$  
& $0.578^{+0.051}_{-0.048}$ 
& $0.602 \pm 0.037$      \\

N$_{BB}$   $(\times 10^{-2})$ 
& $4.49^{+0.78}_{-0.72} $ 
& $4.36^{+0.48}_{-0.45} $ 
& $4.50^{+0.38}_{-0.35} $   
& $4.39^{+0.62}_{-0.53} $  
& $4.10 \pm 0.39 $   
& $3.59^{+0.47}_{-0.40} $   
& $3.97^{+0.38}_{-0.34} $  \\

R$_{BB}$ (km)
& $20 \pm 5$ 
& $20 \pm 3$ 
& $21 \pm 3$ 
& $22 \pm 5$ 
& $22 \pm 4$ 
& $20 \pm 5$ 
& $20 \pm 3$ \\

Flux$_{BB}$  
& $\sim 0.38$ 
& $\sim 0.37$ 
& $\sim 0.38$ 
& $\sim 0.37$ 
& $\sim 0.34$ 
& $\sim 0.30$ 
& $\sim 0.33$ \\
& & & & & & &\\

 kT$_{0}$ (keV) 
& $1.110^{+0.110}_{-0.094}$ 
& $1.102^{+0.066}_{-0.062}$ 
& $1.138 ^{+0.055}_{-0.050}$ 
& $1.065^{+0.084}_{-0.070}$ 
& $1.099^{+0.072}_{-0.068}$ 
& $1.207^{+0.096}_{-0.084}$ 
& $1.285^{+0.073}_{-0.065}$  \\
  
 kT$_{e}$ (keV) 
& $2.710^{+0.060}_{-0.055}$ 
& $2.686 \pm 0.039$  
& $2.643^{+0.042}_{-0.040}$ 
& $2.631^{+0.061}_{-0.057}$ 
& $2.580^{+0.071}_{-0.068}$ 
& $2.592^{+0.116}_{-0.094}$ 
& $2.663^{+0.091}_{-0.082}$   \\

 $\tau$ 
& $12.84 \pm 0.63$ 
& $12.48 \pm 0.40$ 
& $11.86 \pm 0.40$ 
& $12.38 \pm 0.58$ 
& $11.77 ^{+0.70}_{-0.66}$  
& $11.0 \pm 1.1$ 
& $10.82  ^{+0.79}_{-0.83}$  \\

N$_{comptt}$
& $1.023 \pm 0.076$  
& $1.028 \pm 0.050$  
& $1.038 \pm 0.044$  
& $1.100^{+0.071}_{-0.075}$   
& $0.966^{+0.065}_{-0.062}$   
& $1.103^{+0.080}_{-0.088}$     
& $1.237 ^{+0.065}_{-0.072}$  \\

Flux$_{Comptt}$ 
& $\sim 1.56$ 
& $\sim 1.52$ 
& $\sim 1.46$ 
& $\sim 1.54$ 
& $\sim 1.28$ 
& $\sim 1.47$ 
& $\sim 1.76$ \\

R$_W$ (km)
& $6 \pm 1$ 
& $6 \pm 1$ 
& $6 \pm 1$ 
& $7 \pm 1$ 
& $6 \pm 1$ 
& $6 \pm 1$ 
& $5 \pm 1$ \\
& & & & & & &\\
 
Photon Index 
& 2.60 (fixed) 
& 2.60 (fixed) 
& 2.60 (fixed) 
& 2.60 (fixed)  
& 2.60 (fixed)
& 2.60 (fixed)
& 2.60 (fixed)  \\

N$_{po}$     
& $0.19 \pm 0.10$  
& $0.168 \pm 0.061$ 
& $0.139 \pm 0.054$ 
& $0.12 \pm 0.11$ 
& $< 0.044$
& $< 0.44$
& $< 0.071$  \\

Flux$_{po}$     
& $ \sim 0.18$  
& $ \sim 0.16$ 
& $ \sim 0.13$ 
& $ \sim 0.11$ 
& $< 0.04$
& $< 0.41$
& $< 0.07$  \\
& & & & & & &\\

L$_{tot}$ $(\times 10^{37})$ 
& $\sim 4.3$ 
& $\sim 4.1$ 
& $\sim 4.0$ 
& $\sim 4.1$ 
& $<3.4 $  
& $< 4.5$
& $< 4.3$  \\
& & & & & & &\\

$\chi^2$ (d.o.f.)
& 210 (196) 
& 254 (194) 
& 224 (194) 
& 187 (195) 
& 182 (193) 
& 205 (191) 
& 176 (192) \\
\hline     
\hline 
\end{tabular}
\label{tabapofix}
\end{center}
\end{table}

\begin{table}
\begin{center}
\scriptsize 
\caption{ \footnotesize Parameters of the best fit model in the 
  0.12--200 keV  energy band for  spectra A  and B.  The  model is
  defined  in Tab.  \ref{tabnifeperi}. We  fixed the  energies  of the
  absorption edges  present at low energies  finding the corresponding
  optical depths.   Uncertainties are at  90\% confidence level  for a
  single parameter, upper limits are at 95\% confidence level.  Fluxes
  are in units of $10^{-8}$ erg cm$^{-2}$ s$^{-1}$.}
\begin{tabular}{l|c|c} 
\hline
\hline
Parameters 
& A
& B    \\
\hline
$N_{\rm H}$ $\rm (\times 10^{22}\;cm^{-2})$ 
& 0.66 (fixed)
& 0.66 (fixed) \\
& & \\

E$_{\rm O \; {\rm VII}}$ (keV) 
& 0.739 (fixed)
& 0.739 (fixed)\\

$\tau_{\rm O \; {\rm VII}}$ 
& $4.22^{+1.07}_{-0.58}$ 
& $4.17^{+1.10}_{-0.67}$   \\
& & \\

E$_{\rm O \; {\rm VIII}}$ (keV) 
& 0.871 (fixed)
& 0.871 (fixed) \\

$\tau_{\rm O \; {\rm VIII}}$ 
&  $0.76^{+0.43}_{-0.62}$ 
& $<1.08$ \\
& & \\

E$_{\rm Mg \; {\rm XI}}$ (keV) 
& 1.762 (fixed)
& 1.762 (fixed) \\

$\tau_{\rm Mg \; {\rm XI}}$ 
&  $0.103^{+0.090}_{-0.038}$  
&  $0.104 \pm 0.36$    \\
& & \\

E$_{\rm Mg \; {\rm XII}}$ (keV)
& 1.963 (fixed)
& 1.963 (fixed) \\
 
$\tau_{\rm Mg \; {\rm XII}}$ 
&  $0.052^{+0.059}_{-0.017}$  
&  $0.048 \pm 0.017$    \\
& & \\

E$_{\rm Fe \; {\rm I}}$ (keV) 
& 7.117 (fixed) 
& 7.117 (fixed) \\

$\tau_{\rm Fe \; {\rm I}}$ 
& $<0.094$
& $<0.089$  \\
& & \\

E$_{edge_1}$ (keV)
& $8.447 \pm 0.057$
& $8.464 \pm 0.041$   \\

$\tau_{edge_1}$ 
& $0.84^{+0.12}_{-0.10}$ 
& $0.732 ^{+0.093}_{-0.080}$    \\
& & \\

E$_{\rm O\;{\rm VIII}}$ (keV)
& 0.6537 (fixed)
& 0.6537 (fixed)\\

$\sigma_{\rm O\;{\rm VIII}}$ (keV)
&  $<0.34$
&  $<0.06$ \\

I$_{\rm O\;{\rm VIII}}$    
& $-1.36 \pm 0.27$
& $-1.24^{+0.29}_{-0.35}$\\

EQW$_{\rm O\;{\rm VIII}}$ (eV)
&$-125 \pm 25$ 
&$-135^{+33}_{-37}$\\
& & \\

E$_{\rm Mg\;{\rm XII}}$ (keV)
& 1.4726 (fixed)
& 1.4726 (fixed)\\

$\sigma_{\rm Mg\;{\rm XII}}$ (keV)
&  $<0.18$
&  $<0.08$ \\

I$_{\rm Mg\;{\rm XII}}$
& $-0.171^{+0.075}_{-0.610}$
& $-0.140^{+0.050}_{-0.097}$\\

EQW$_{\rm Mg\;{\rm XII}}$ (eV)
&$-19^{+8}_{-69}$
&$-17^{+6}_{-12}$  \\
& & \\

E$_{\rm Fe\;{\rm XXIII}}$ (keV)
& 6.62 (fixed)
& 6.62 (fixed)\\

$\sigma_{\rm Fe\;{\rm XXIII}}$ (keV)
&  $<0.31$
&  0.31 (fixed) \\

I$_{\rm Fe\;{\rm XXIII}}$  $(\times 10^{-3})$ 
& $1.12^{+0.64}_{-0.74}$
& $1.75^{+0.81}_{-0.89}$\\

EQW$_{\rm Fe\;{\rm XXIII}}$ (eV)
&$29^{+18}_{-20}$
&$39 \pm 20$  \\
& & \\

 kT$_{0}$ (keV) 
& $0.430^{+0.015}_{-0.035}$  
& $0.447 \pm 0.014$ \\
  
 kT$_{e}$ (keV) 
& $0.959^{+0.083}_{-0.052}$ 
& $1.018^{+0.080}_{-0.054}$ \\

 $\tau$ 
& $13.5^{+1.2}_{-1.6}$  
& $13.2^{+1.9}_{-3.0}$  \\

N$_{comptt}$
& $29.5^{+2.8}_{-3.9}$ 
& $25.9^{+1.9}_{-3.0}$   \\

Flux$_{Comptt}$ 
& $\sim 4.9$ 
& $\sim 4.8$ \\

R$_W$ (km)
& $100 \pm 20$ 
& $90 \pm 10$ \\
& & \\
 
Photon Index 
& 2.60 (fixed)
& 2.60 (fixed) \\

N$_{po}$     
& $0.322 \pm 0.065$ 
& $0.219 ^{+0.055}_{-0.064}$   \\

Flux$_{po}$     
& $ \sim 0.30$  
& $ \sim 0.21$ \\
& & \\
L$_{tot}$ $(\times 10^{38})$ 
& $\sim 1.0$
& $\sim 1.0$  \\
 & &\\

$\chi^2$ (d.o.f.)
& 187 (170) 
& 207 (170) \\
\hline     
\hline 
\end{tabular}
\label{tabperifix}
\end{center}
\end{table}

\begin{table*}
\begin{center}
\caption { In column 1 the radial velocities  $v_r$ corresponding 
  to the  absorption features present in  the HI spectrum  of Cir X--1
  are reported  (see Goss \& Mebold  1977); in column 2  the radius of
  the orbit of  the clouds along the line of sight.   In columns 3 the
  corresponding distances to the clouds of neutral matter.  In columns
  4  and 5  the visual  extinction $A_v$  and the  equivalent hydrogen
  column N$_H$ corresponding to the  distances; in column 6 the proper
  motion  of  the clouds.   These  values  are  calculated assuming  a
  distance  to the  Galactic center  of $8.5  \pm 0.5$  kpc  (Feast \&
  Whitelock, 1997). The  velocity at 3 km/s gives  only one value
  of $d$  because its orbit  has a radius  larger than 8.5  kpc.  (see
  text). }
\begin{tabular}{c|c|c|c|c|c} 
\hline
\hline 
$v_r$   & R   & d  & $A_{v}$  & N$_{H}$   & $\mu$ \\
km/s    &kpc  &kpc & mag      & cm$^{-2}$ &mas/yrs\\
  
\hline 

-75.0& $5.3 \pm 0.3$& $5.9 \pm 0.8$  &$6.4^{+3.0}_{-2.5}$&$1.14 \pm 0.21$&
$0.4 \pm 0.5$ \\
       &       &$7.6 \pm 0.8$  &$8.9^{+3.9}_{-3.2}$  &$1.59 \pm 0.24$&
$-0.3 \pm 0.3$ \\

 & & & & &  \\
-57.5 &$5.8 \pm 0.4$&$4.1 \pm 0.3$ & $3.7 \pm 1.4$& $0.66 \pm 0.09$& 
 $1.4 \pm 0.3$ \\

       & &$9.3 \pm 0.6$ & $11.4^{+4.3}_{-3.7}$& $2.04 \pm 0.20$& 
 $-0.6 \pm 0.2$ \\
 & & & & &  \\

-21 &$7.8 \pm 0.5$&$0.9 \pm 0.1$ & $0.73 \pm 0.35$& $0.130 \pm 0.005$& 
 $5.5 \pm 1.0$ \\

  & & $12.5 \pm 0.7$ & $16.1^{+6.4}_{-5.5}$& $2.88 \pm 0.23$& 
 $-0.39 \pm 0.06$ \\
 & & & & &  \\

-6.5 &$8.1 \pm 0.5$&$0.5 \pm 0.1$ & $0.50^{+0.35}_{-0.30}$& $0.090 \pm 0.016$& 
 $3.3 \pm 1.0$ \\

 &   &$12.9 \pm 0.8$ & $16.7^{+6.9}_{-5.9}$& $2.98 \pm 0.26$& 
 $-0.13 \pm 0.02$ \\
 & & & & &  \\

   3.0  & $8.6 \pm 0.5$ &$13.6 \pm 0.8$&$17.7^{+7.6}_{-6.3}$ &$3.17 \pm 0.27$&
 $0.06 \pm 0.02$  \\
 
\hline 
\hline 
\end{tabular}
\label{tabdistances}
\end{center}
\end{table*}

\begin{table}
\scriptsize 
\begin{center}
\caption{Summary of the distance $d$, the electron densities $n_e$ and
the equivalent hydrogen column $N$ associated to the ionized elements.
$f$ and Log$_{10} \xi$ are defined in the text.}
\begin{tabular}{l|c|c|c|c|c|c|c} 
\hline
\hline
 Phases 
& O {\sc vii} 
& O {\sc viii}    
& O {\sc viii}  
& Ne {\sc ix} 
& Mg {\sc xi} 
& Mg {\sc xi} 
& Mg {\sc xii}  \\

0.62--0.84
& edge
& edge
&line
&edge
&edge
&edge
&line\\

 & & & & & & &  \\

Log$_{10} \xi$
& 1.3
& 1.3
& --
& 1.3
& --
& --
& --\\

$f$
&0.5
&0.5
&--
&0.5
& --
& --
& --\\

d (cm)
&  $<5.3 \times 10^{13}$  
& $<5.3 \times 10^{13}$
& --
& $<5.3 \times 10^{13}$
& --
& --
& --\\

N (cm$^{-2}$)
&  $\sim 3 \times 10^{22}$
&  $\sim 4 \times 10^{22}$
& --
&  $\sim 2 \times 10^{22}$
& --
& --
& --\\

n$_e$ (cm$^{-3}$)
& --
& --
& --
& --
& --
& --
& --\\
\hline

Phase & & & & & &  \\
0.11--0.16 & & & & & & \\

& & & & & & &  \\

Log$_{10} \xi$
& 1.0
& 1.0
& 1.0
& --
& 2.2
& 2.2
& 2.2\\

$f$
& 0.5
& 0.25
& 0.25
& --
& 0.4
& 0.45
& 0.45\\

d (cm)
& $<2.5 \times 10^{14}$
& $<2.5 \times 10^{14}$
& $ \sim 1 \times 10^{13}$
& --
& $<1 \times 10^{13}$ 
& $<1 \times 10^{13}$
& $ \sim 1 \times 10^{11}$\\

N (cm$^{-2}$)
& $\sim 4 \times 10^{22}$
& $\sim 4 \times 10^{22}$
& --
& --
& $\sim 7 \times 10^{22}$
& $\sim 7 \times 10^{22}$
& --\\

n$_e$ (cm$^{-3}$)
& --
& --
& $ \sim 9.2 \times 10^{10}$
& --
& --
& --
& $ \sim 6.2 \times 10^{13}$\\

\hline     
\hline 
\end{tabular}

\vspace{1. cm}
\begin{tabular}{l|c|c|c} 
\hline
\hline
 Phases 
&Fe {\sc xxiii} 
&Fe {\sc xxv}  
&Fe {\sc xxv} \\

0.62--0.84
&edge
&line
&edge\\

  & & & \\

Log$_{10} \xi$
& --
& 3.3
& 3.3 \\

$f$
& --
& 0.65
& 0.65 \\

d (cm)
& --
& $ \sim 2\times 10^{10}$
& $<2.5 \times 10^{11}$ \\

N (cm$^{-2}$)
& --
& --
&  $\sim 8.3 \times 10^{22}$ \\

n$_e$ (cm$^{-3}$)
& --
& $n_e \sim 7 \times 10^{13}$
& -- \\
\hline

Phase  & & &  \\
0.11--0.16 & & &  \\

& & &   \\

Log$_{10} \xi$
& 2.9
& --
& -- \\

$f$
& 0.25
& --
& -- \\

d (cm)
& $<3.3 \times 10^{10}$
& --
& -- \\

N (cm$^{-2}$)
& $\sim 4 \times 10^{24}$
& --
& -- \\

n$_e$ (cm$^{-3}$)
& --
& --
& -- \\
\hline     
\hline 
\end{tabular}
\label{tabioni}
\end{center}
\end{table}

\clearpage

\end{document}